\documentclass[aps,prd,reprint,showkeys,nofootinbib,superscriptaddress,notitlepage]{revtex4-1}
\usepackage{amsmath,graphicx,amssymb,multirow,url,xspace,natbib}

\setcitestyle{authoryear,open={(},close={)}}

\usepackage[utf8]{inputenc}
\usepackage[unicode=true,
  bookmarks=false,   backref=false,    colorlinks=true,
  linktocpage=true,  citecolor=black,  linkcolor=black,
  urlcolor=black,    breaklinks=true
]{hyperref}

\newcommand{\Rbase}{R22a\xspace}
\newcommand{\RGaia}{R21\xspace}

\newcommand{\MHW}{M_{H,1}^W}

\newcommand{\be}{\begin{equation}}
\newcommand{\ee}{\end{equation}}

\newcommand{\sy}{\scriptstyle}

\newcommand{\td}{..}
\newcommand{\oh}{{\rm [O/H]}}
\newcommand{\p}{{[P]}}

\begin{document}

\title{
Reassessing the Cepheid-based distance ladder: implications for the Hubble constant
}

\author{Marcus \surname{Högås}}
\email{marcus.hogas@fysik.su.se}
\affiliation{Oskar Klein Centre, Department of Physics, Stockholm University\\Albanova University Center\\ 106 91 Stockholm, Sweden}

\author{Edvard \surname{Mörtsell}}
\email{edvard@fysik.su.se}
\affiliation{Oskar Klein Centre, Department of Physics, Stockholm University\\Albanova University Center\\ 106 91 Stockholm, Sweden}

\begin{abstract}
The Hubble constant ($H_0$) is a key parameter in cosmology, yet its precise value remains contentious due to discrepancies between early- and late-universe measurement methods, a problem known as the ``Hubble tension.'' In this study, we revisit the Cepheid-based distance ladder calibration, focusing on two potential sources of bias in the period-luminosity relation (PLR): (1) the assumed prior for the residual parallax offset of the Milky Way (MW) Cepheids and (2) systematic differences between Cepheid periods in anchor galaxies versus supernova host galaxies.
To address the latter, we adopt two different strategies alongside a renewed MW Cepheid calibration. The first strategy involves resampling anchor and host Cepheids from a common distribution of periods. This approach provides a conservative estimate of $H_0 = (72.18 \pm 1.76) \, \mathrm{km/s/Mpc}$, including the renewed MW analysis. 
The increased uncertainty reflects the reduced sample size---about 700 Cepheids per resampling compared to 3200 in the original dataset.
This method reduces the Hubble tension from $5.4 \, \sigma$ (as reported by the SH0ES collaboration with $H_0 = (73.17 \pm 0.86) \, \mathrm{km/s/Mpc}$) to $2.4 \, \sigma$.
The second strategy allows the PLR slope to vary with the period, yielding $H_0 = (72.35 \pm 0.91) \, \mathrm{km/s/Mpc}$, including the renewed MW analysis, and the tension reduced to $4.4 \, \sigma$.
A statistical comparison of the model with the single-linear PLR shows a significant preference for the broken PLR (p-value $< 0.001$).
Both strategies consistently indicate a downward shift of approximately $-1 \, \mathrm{km/s/Mpc}$ in $H_0$.
Our findings underscore the importance of careful consideration of Cepheid population characteristics for precise $H_0$ calibrations.
\end{abstract}

\begin{keywords}
{stars: variables: Cepheids -- cosmology: cosmological parameters -- cosmology: distance scale -- galaxies: distances and redshifts}
\end{keywords}

\maketitle

\section{Introduction}
The measurement of the Hubble constant, $H_0$, the rate at which the universe is expanding, is a cornerstone of modern cosmology. Since its first estimation by Edwin Hubble in 1929 \citep{Hubble:1929}, it has enabled the scale and age of the Universe to be determined.
However, the exact value of $H_0$ has remained contentious, with recent advances in observational techniques exacerbating the discrepancies between methods. This discrepancy, known as the “Hubble tension,” has emerged as one of the most significant puzzles in cosmology, particularly between the two main avenues of measurement: the early-universe model-dependent methods (extrapolated from cosmic microwave background radiation) and the late-universe model-independent methods, where the Cepheid-based distance ladder plays a central role.

The calibration of the Cepheid-based distance ladder has undergone numerous refinements. In recent years, the SH0ES (Supernova H0 for the Equation of State) team’s work has been instrumental, providing a local-universe measurement of $H_0$ through an intricate calibration of Cepheids in anchor galaxies, including the Large Magellanic Cloud (LMC), the Small Magellanic Cloud (SMC), the Milky Way (MW), and NGC 4258. Their results for $H_0$ have been reported in the range around $73-74 \, \mathrm{km/s/Mpc}$ \citep{Riess:2009pu,Riess:2011yx,Riess:2016jrr,Riess_2018,Riess:2018byc,Riess_2019,Riess_2021,Riess:2021jrx,Riess:2022mme,Murakami:2023xuy,Breuval:2024lsv} with their most updated value $H_0 = (73.17 \pm 0.86) \, \mathrm{km/s/Mpc}$ \citep{Breuval:2024lsv} being $5.4 \, \sigma$ higher than $H_0 = (67.8 \pm 0.5) \, \mathrm{km/s/Mpc}$, derived from early-universe measurements using the Planck satellite, assuming a $\Lambda$CDM model \citep{Planck:2018vyg}.

In this work, we explore two potential sources of unaccounted-for systematic effects in the Cepheid-based distance ladder.
The first is reassessing the calibration of the MW Cepheids, in particular the choice of prior for the residual parallax offset, which leads to a $-0.6 \, \mathrm{km/s/Mpc}$ shift of the Hubble constant, compared with the standard SH0ES calibration, or $-1.4 \, \mathrm{km/s/Mpc}$ if the MW is used as the sole anchor. 

The second is the difference in the distribution of Cepheid periods between anchor and supernova host galaxies.
The Cepheids in the anchor galaxies calibrate the zero-point of the period-luminosity relation (PLR), and those in host galaxies determine the distances to the type Ia supernovae.
Specifically, we find that Cepheids with longer periods, predominantly in host galaxies, tend to yield higher $H_0$ values, suggesting a possible systematic offset in the calibration.
To mitigate this effect, we use two different approaches:
\begin{enumerate}
    \item A matched-period resampling method, harmonizing the distribution of periods in the anchor and host galaxies. 
    This leads to an additional $-0.5 \, \mathrm{km/s/Mpc}$ shift in $H_0$. 
    The method also increases the uncertainty, from $0.9 \, \mathrm{km/s/Mpc}$ to $1.8 \, \mathrm{km/s/Mpc}$, due to a reduced sample size. 
    In conjunction with the renewed MW analysis, this leads to $H_0 = (72.18 \pm 1.76) \, \mathrm{km/s/Mpc}$, to be compared with the SH0ES team's $H_0 = (73.17 \pm 0.86) \, \mathrm{km/s/Mpc}$. Thus, there is a reduction in the Hubble tension from $5.4 \, \sigma$ to $2.4 \, \sigma$.
    \item Another potential way of mitigating a bias due to the difference in the Cepheid periods is to generalize the single-linear PLR, allowing for different slopes in different period ranges. This is referred to as the broken PLR model.
    The data provides strong evidence against the single-linear PLR, favoring the broken PLR with a confidence level of $> 99.9 \, \%$.
    The broken PLR gives a $-0.3 \, \mathrm{km/s/Mpc}$ shift in $H_0$, thus pointing in the same direction as the resampling method.
    Since all Cepheids are retained in the generalized PLR fit, the uncertainty in $H_0$ is roughly the same as that of the SH0ES team analysis.
    In conjunction with the renewed MW analysis, this method yields $H_0 = (72.35 \pm 0.91) \, \mathrm{km/s/Mpc}$ corresponding to a reduction in the tension from $5.4 \, \sigma$ to $4.4 \, \sigma$. 
\end{enumerate}
\begin{figure}[t]
    \centering
    
    \includegraphics[width=\linewidth]{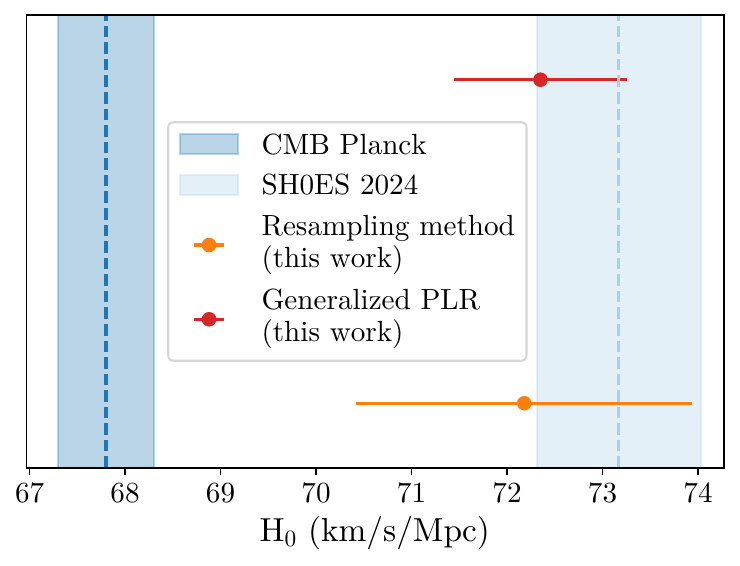}
    
    \caption{Key results. The left shaded region represents the early-universe $H_0$ estimate based on data from the Planck satellite, assuming a $\Lambda$CDM model \citep{Planck:2018vyg} and the right shaded region the most updated Cepheid-based local distance ladder value from the SH0ES team \citep{Breuval:2024lsv}. There is a $5.4 \, \sigma$ tension between the two.
    The orange data point with error bars shows the Cepheid-based distance ladder result when using a resampling method to mitigate the period discrepancy between the anchor and host Cepheids, together with a renewed analysis of the Milky Way Cepheids. In this case, the tension is reduced to $2.4 \, \sigma$.
    The red data point with error bars shows the same when the single-linear period-luminosity relation is replaced with a PLR allowing for different slopes in different period ranges. In this case, the tension is reduced to $4.4 \, \sigma$.
    \label{fig:H0_summary}}
\end{figure}
These findings are summarized in Fig.~\ref{fig:H0_summary} and emphasize the need for careful consideration of Cepheid properties in resolving the ongoing Hubble tension.

For the readers wanting to make their own recalibration of the Cepheid-based distance ladder, we have invested some effort into presenting the high-level SH0ES data in \cite{Riess:2021jrx} in a transparent fashion.\footnote{Publicly available at \href{https://github.com/marcushogas/Cepheid-Distance-Ladder-Data}{\url{https://github.com/marcushogas/Cepheid-Distance-Ladder-Data}}}\\

\noindent \textbf{Notation.} Galaxy names: N4258 stands for NGC 4258 and U9391 stands for UGC 9391, etc. Throughout, $\log = \log_{10}$.


\section{Data description}
We adopt the data presented in the latest (i.e., fourth) iteration of SH0ES \citep{Riess:2021jrx}, hereafter \Rbase, primarily via the publicly available fits files.\footnote{\href{https://github.com/PantheonPlusSH0ES/DataRelease/tree/main/SH0ES_Data}{\url{https://github.com/PantheonPlusSH0ES/DataRelease/tree/main/SH0ES_Data}} (last checked 2024-03-27)} We also include all recent updates, with data available in \cite{Riess:2022mme,Bhardwaj:2023mau,Murakami:2023xuy,Breuval:2024lsv}. 
More details about the exact data usage is presented continuously.

\section{The Cepheid-based distance ladder}
\subsection{First rung}
As discovered by Leavitt and Pickering \citep{Leavitt:1907,Leavitt:1912zz}, Cepheid variable stars exhibit a well-defined period-luminosity relation which allows them to be used as standardizable candles for measuring astronomical distances.
By the first rung of the distance ladder, we mean Cepheids in galaxies with direct distance measurements, referred to as anchor galaxies.\footnote{An additional requirement is for the Cepheids to exhibit HST photometry.}

Currently, distance ladder calibrations, such as \Rbase, typically use redenning-free magnitudes, also called ``Wesenheit'' magnitudes.
We adopt the following definition of the Wesenheit magnitude,
\begin{equation}
    \label{eq:mHW_def}
    m_H^W = m_{H} - R (m_V-m_I),
\end{equation}
where $m_X$ is the observed $X$-band apparent magnitude ($X = H,V,I$) and $R$ is the total-to-selective extinction coefficient, correcting for dust extinction and, to some degree, also for the intrinsic color-luminosity dependence see \cite{Madore:1991yf}.
As such, the value for $R$ may differ between Cepheids in different galaxies. However, the \Rbase baseline analysis sets $R = 0.386$ throughout, that is, assuming Milky Way-like reddening in all galaxies. Since we are using the \Rbase-values for $m_H^W$ in our analysis, we are also implicitly assuming this reddening law. See \cite{Mortsell:2021tcx,mortsell2021hubble} for alternative analyses.

The PLR can be derived from relatively simple physical considerations using the Stefan--Boltzmann law and a well-known relation for mechanical systems, $P \propto 1 / \sqrt{\rho}$, where $P$ is the period of the Cepheid and $\rho$ is the density \citep{Madore:1991yf}.
The result is,
\begin{equation}
    \label{eq:PLR}
    m_{H,i}^W = \mu_i + \MHW + b_W \p_i + Z_W \oh_i.
\end{equation}
Here, $\mu_i$ is the distance modulus of Cepheid number $i$,  and $\oh_i$ is its metallicity, quantified as $\oh_i = \log \left[ (\mathrm{O/H})_i / (\mathrm{O/H})_\odot \right]$.
Moreover, $\p_i = \log P_i / \mathrm{days} - 1$.
In eq.~\eqref{eq:PLR}, we have allowed for a possible metallicity dependence, thus ``period-luminosity-color-metallicity relation'' would be an appropriate name, although we will refer to it simply as the period-luminosity relation. 
In eq.~\eqref{eq:PLR}, $\MHW$, $b_W$, and $Z_W$ are empirical constants, common for all Cepheids. These constants determine the intercept and slopes of the relation.

In anchor galaxies, where a geometric distance estimate for $\mu$ can be provided, the PLR \eqref{eq:PLR} can be calibrated from the Cepheid photometry. That is, provided measurements of the Cepheid magnitudes, colors, periods, and metallicity, the empirical constants $M_{H,1}^W$, $b_W$, and $Z_W$ can be determined.

\subsection{Second rung}
The second rung of the distance ladder consists of standardized Cepheids in SNIa host galaxies. We refer to these type Ia supernovae as ``Cal SNe'' (Cepheid Calibrator SNe).
With the PLR being determined in the first rung, the distances to these galaxies can be established based on measurements of the Cepheids' apparent magnitude, period, and metallicity. 

The standardized peak apparent magnitude for SNIa number $i$ (in the $B$-band) is given by
\begin{equation}
    m_{B,i}= \mu_i + M_B 
\end{equation}
where $M_B$ is the fiducial SNIa absolute magnitude. The SNIa standardization consists in correcting the magnitude for the SNIa color, width of the light curve, and host galaxy dependence, see \cite{Scolnic:2021amr}.
With $m_{B,i}$ being measured and $\mu_i$ determined by the Cepheids, the fiducial SNIa absolute magnitude $M_B$ can be constrained.

\subsection{Third rung}
The third and final rung of the Cepheid-based distance ladder consists of SNIa in the Hubble flow (HF SNe) where the expansion of the Universe gives the dominant contribution to the redshift.
A Taylor expansion around redshift $z=0$ for HF SNIa number $i$ yields,
\begin{equation}
    \label{eq:mBHFSN}
    m_{B,i} - 5 \log c z_i \left\lbrace 1 + \frac{1}{2} (1-q_0) z_i \right\rbrace - 25 = M_B - 5 \log H_0
\end{equation}
to second order in redshift where $q_0$ is the deceleration parameter. 
Since $M_B$ is calibrated in the second rung, the Hubble constant $H_0$ and deceleration parameter can be inferred from observations of the HF SNe magnitudes and redshift. Alternatively, $q_0$ can be provided from external sources as done in the \Rbase baseline analysis where it is set to $q_0 = -0.55$. Since we adopt the values for the left-hand side of eq.~\eqref{eq:mBHFSN} from \Rbase, we are also implicitly assuming this value. 

Finally, note that although we have presented the three rungs separately, \Rbase makes a simultaneous calibration of all three rungs (except for the MW Cepheids, as we discuss below).

\section{The 2022 SH0ES-team calibration}
\label{sec:R22}
The methodology we present here is equivalent to the \Rbase baseline calibration although with a few formal differences that we have introduced for clarity of the presentation. This includes a few reparameterizations and reorderings of the data. The final value of the inferred model parameters in this section are identical with those of \Rbase.

The approach of \Rbase allows for a linear fit of the distance ladder.
The data points are placed in the data vector $y$, and the free parameters in the vector $q$, the design matrix is denoted $L$, and the covariance matrix is denoted $C$. Thus, $Lq$ gives the model prediction for $y$ and the $\chi^2$ statistic reads
\begin{equation}
    \label{eq:chi2_R22}
    \chi^2 = \left( y - Lq \right)^T C^{-1} \left( y - Lq\right).
\end{equation}
The \Rbase analysis exhibits the following form for these quantities:
\begin{widetext}
    \begin{equation}
    \label{eq:y_R22}
    y = 
    \begin{array}{ll}
    \left(
    \begin{array}[c]{c}
    
    m^W_{H,\mathrm{M101}} \\

    : \\

    m^W_{H,\mathrm{U9391}} \\
    
    \hline
    
    m^W_{H,\textrm{N4258}} \\
    
    m^W_{H,\textrm{M31}} \\
    
    m^W_{H,\textrm{LMC,GRND}} \\

    m^W_{H,\textrm{LMC,HST}} \\
    
    m^W_{H,\textrm{SMC}} \\
    
    \hline
    
    m_{B,\mathrm{Cal \; SN}} \\

    \hline
    
    m_{B,\mathrm{HF \; SN}} - 5 \log c z \{ ... \} -25 \\
    
    \hline
    
    M_{H,1,\textrm{HST}}^W \\
    
    M_{H,1,{\textrm{Gaia}}}^W \\
    
    Z_{W} \\
    
    0 \\
    
    \mu_\mathrm{N4258}^\mathrm{anch} \\
    
    \mu_\mathrm{LMC}^\mathrm{anch}
    \end{array} \right)
    
    &
    
    \begin{array}[c]{@{}l@{\,}l}
    
    \left.
    \begin{array}{c} \vphantom{m^W_{H,\mathrm{hosts}}} \\
    \vphantom{:} \\
    \vphantom{m^W_{H,\mathrm{hosts}}}
    \end{array}
    \right\} & \text{2150 Cepheids in SNIa hosts} \\
    
    \left.
    \begin{array}{c} \vphantom{m^W_{H,\textrm{nh},j}} \\ 
    \vphantom{m^W_{H,\textrm{nh},j}} \\ 
    \vphantom{m^W_{H,\textrm{nh},j}} \\
    \vphantom{m^W_{H,\textrm{nh},j}} \\
    \vphantom{\textrm{\LARGE HELLO}}
    \end{array}
    \right\} & \text{(443 + 55 + 270 + 69 + 143) Cepheids in anchors or non-SNIa hosts\hspace{1.5in}} \\
    
    \left.
    \begin{array}{c}
    \vphantom{m_B^0}
    \end{array}
    \right\} & \text{77 Cal SNe magnitudes} \\

    \left.
    \begin{array}{c}
    \vphantom{.}
    \end{array}
    \right\} & \text{277 HF SNe} \\
    
    \left.
    \begin{array}{c}
    \vphantom{.} \\
    \vphantom{.} \\
    \vphantom{.} \\
    \vphantom{.} \\ 
    \vphantom{.} \\
    \vphantom{.}
    \end{array}
    \right\} & \text{6 External constraints}
    \end{array}
    \end{array}
    \end{equation}

    \begin{equation}
    \label{eq:C_R22}
        C = \left(
        \begin{array}{cccccccccccccccc}
        
        \sy{\sigma_{\rm M101}^2}\!\!\!\! & \td & \sy{Z_{\textrm{cov}}} & \sy{Z_{\textrm{cov}}} & \sy{0} & \sy{0} & \sy{0} & \sy{0} & \sy{0} & \sy{0}  & \sy{0} & \sy{0} & \sy{0} & \sy{0} & \sy{0} & \sy{0} \\
        
        : & \rotatebox{45}{:} & : & : & : & : & : & : & : & : & : & : & : & : & : & : \\
        
        \sy{Z_{\textrm{cov}}} & \td & \sy{\sigma_{\rm U9391}^2}\!\!\!\! & \sy{Z_{\textrm{cov}}} & \sy{0} & \sy{0} & \sy{0} & \sy{0} &\sy{0} & \sy{0} & \sy{0} & \sy{0} & \sy{0} & \sy{0}  & \sy{0} & \sy{0} \\
        
        \hline
        
        \sy{Z_{\textrm{cov}}} & \td & \sy{Z_{\textrm{cov}}} & \sy{\sigma_{{\rm N4258}}^2}\!\!\!\! & \sy{0} & \sy{0} & \sy{0} & \sy{0}  &\sy{0} & \sy{0} & \sy{0} & \sy{0} & \sy{0} & \sy{0}  & \sy{0} & \sy{0} \\
        
        \sy{0} & \td & \sy{0} & \sy{0} & \sy{\sigma_{{\rm M31}}^2}\!\!\!\! & \sy{0} & \sy{0} & \sy{0}  &\sy{0} & \sy{0} & \sy{0} & \sy{0} & \sy{0} & \sy{0}  & \sy{0} & \sy{0} \\

        \sy{0} & \td & \sy{0} & \sy{0} & \sy{0} &  \sy{\sigma_{{\rm LMC,GRND}}^2}\!\!\!\! & \sy{10^{-4}} & \sy{0} & \sy{0} & \sy{0} & \sy{0} & \sy{0} & \sy{0} & \sy{0}  & \sy{0} & \sy{0} \\

        \sy{0} & \td & \sy{0} & \sy{0} & \sy{0} & \sy{10^{-4}} &  \sy{\sigma_{{\rm LMC,HST}}^2}\!\!\!\! & \sy{0} & \sy{0} & \sy{0} & \sy{0} & \sy{0} & \sy{0} & \sy{0} & \sy{0}  & \sy{0} \\

        \sy{0} & \td & \sy{0} & \sy{0} & \sy{0} & \sy{0} & \sy{0} &  \sy{\sigma_{{\rm SMC}}^2}\!\!\!\! & \sy{0} & \sy{0} & \sy{0} & \sy{0} & \sy{0} & \sy{0} & \sy{0}  & \sy{0} \\

        \hline
                
        \sy{0} & \td & \sy{0} & \sy{0} & \sy{0} & \sy{0} & \sy{0} & \sy{0} & \sy{\sigma^2_{\textrm{Cal SN}}}\!\!\!\! & \sy{{\rm SN}_{\textrm{cov}}} & \sy{0} & \sy{0} & \sy{0} & \sy{0} & \sy{0} & \sy{0} \\
        
        \hline

        \sy{0} & \td & \sy{0} & \sy{0} & \sy{0} & \sy{0} & \sy{0} & \sy{0} & \sy{{\rm SN}_{\textrm{cov}}} & \sy{\sigma^2_{\textsc{HF SN}}}\!\!\!\! & \sy{0} & \sy{0} & \sy{0} & \sy{0} & \sy{0} & \sy{0} \\
        
        \hline
        
        \sy{0} & \td & \sy{0} & \sy{0} & \sy{0} & \sy{0} & \sy{0} & \sy{0} & \sy{0} & \sy{0} & \sy{\sigma_{\rm HST}^2} & \sy{0}\!\!\!\! & \sy{0} & \sy{0} & \sy{0}  & \sy{0} \\
        
        \sy{0} & \td & \sy{0} & \sy{0} & \sy{0} & \sy{0} & \sy{0} & \sy{0} & \sy{0} &  \sy{0} & \sy{0} & \sy{\sigma_{\rm Gaia}^2} & \sy{0}\!\!\!\! & \sy{0} & \sy{0} & \sy{0} \\

        \sy{0} & \td & \sy{0} & \sy{0} & \sy{0} & \sy{0} & \sy{0} & \sy{0} & \sy{0} &  \sy{0} & \sy{0} & \sy{0} & \sy{\sigma_{Z_W}^2} & \sy{0}\!\!\!\! & \sy{0} & \sy{0} \\
        
        \sy{0} & \td & \sy{0} & \sy{0} & \sy{0} & \sy{0} & \sy{0} & \sy{0} & \sy{0} & \sy{0} & \sy{0} & \sy{0} & \sy{0} & \sy{\sigma_{\rm grnd}^2}\!\!\!\! & \sy{0} & \sy{0} \\
        
        \sy{0} & \td & \sy{0} & \sy{0} & \sy{0} & \sy{0} & \sy{0} & \sy{0} & \sy{0} & \sy{0} & \sy{0} & \sy{0} & \sy{0} & \sy{0} & \sy{\sigma_{\mu,{\rm N4258}}^2}\!\!\!\! & \sy{0} \\
        
        \sy{0} & \td & \sy{0} & \sy{0} & \sy{0} & \sy{0} & \sy{0} & \sy{0} &  \sy{0} & \sy{0} & \sy{0} & \sy{0} & \sy{0} & \sy{0} & \sy{0} & \sy{\sigma_{\mu,{\rm LMC}}^2}
        \end{array}
        \right)
    \end{equation}

    \begin{equation}
    \label{eq:L_q_R22}
        L =
        \left(
        \begin{array}[c]{ccccccclclcc}
        
        1 & \td & 0 & 0 & 1 & 0 & 0 & [P]_\mathrm{M101} & 0 & \oh_\mathrm{M101} & 0 & 0 \\
        
        : & \rotatebox{45}{:} & : & : & : & : & :   & : & : & : & : & : \\  
        
        0 & \td & 1 & 0 & 1 & 0 & 0 & [P]_\mathrm{U9391} & 0 & \oh_\mathrm{U9391} & 0 & 0 \\
        
        \hline
        
        0 & \td & 0 & 1 & 1 & 0 & 0 & [P]_\mathrm{N4258} & 0 & \oh_{\textrm{N4258}} & 0 & 0 \\
        
        0 &\td & 0 & 0 & 1 & 0 & 1 & [P]_\mathrm{M31} & 0 & \oh_{\textrm{M31}}   & 0 & 0 \\
        
        0 & \td & 0 & 0 & 1 & 1 & 0 & [P]_\mathrm{LMC,GRND} & 0 & \oh_{\textrm{LMC,GRND}} & 1 & 0 \\

        0 & \td & 0 & 0 & 1 & 1 & 0 & [P]_\mathrm{LMC,HST} & 0 & \oh_{\textrm{LMC,HST}} & 0 & 0 \\

        0 & \td & 0 & 0 & 1 & 1 & 0 & [P]_\mathrm{SMC} & 0 & \oh_{\textrm{SMC}}    & 1 & 0 \\
        
        \hline
        
        1 & \td & 0 & 0 & 0 & 0 & 0 & 0 & 1 & 0 & 0 & 0 \\
        
        : & \rotatebox{45}{:} & : & : & : & : & : & : & : & : & : & : \\  
        
        0 & \td & 1 & 0 & 0 & 0 & 0 & 0 & 1 & 0 & 0 & 0 \\
        
        \hline

        0 & \td & 0 & 0 & 0 & 0 & 0 & 0 & 1 & 0 & 0 & -1 \\

        : & \rotatebox{45}{:} & : & : & : & : & : & : & : & : & : & : \\
        
        0 & \td & 0 & 0 & 0 & 0 & 0 & 0 & 1 & 0 & 0 & -1\\

        \hline
        
        0 & \td & 0 & 0 & 1 & 0 & 0 & 0 & 0 & 0 & 0 & 0 \\
        
        0 & \td & 0 & 0 & 1 & 0 & 0 & 0 & 0 & 0 & 0 & 0 \\

        0 & \td & 0 & 0 & 0 & 0 & 0 & 0 & 0 & 1 & 0 & 0 \\
        
        0 & \td & 0 & 0 & 0 & 0 & 0 & 0 & 0 & 0 & 1 & 0 \\
        
        0 & \td & 0 & 1 & 0 & 0 & 0 & 0 & 0 & 0 & 0 & 0 \\
        
        0 & \td & 0 & 0 & 0 & 1 & 0 & 0 & 0 & 0 & 0 & 0
        
        \end{array}
        \right), \quad \quad
        q = 
        \left(
        \begin{array}{c}
            \mu_\mathrm{M101} \\
            : \\
            \mu_\mathrm{U9391} \\
            \hline
            \mu_\mathrm{N4258} \\
            M_{H,1}^W \\
            \mu_\mathrm{LMC} \\
            \mu_\mathrm{M31} \\
            b_W \\
            M_B \\
            Z_W \\
            \Delta \mathrm{zp} \\
            5 \log H_0
        \end{array}
        \right) .
    \end{equation}
\end{widetext}
These are publicly available, in this form, on GitHub.\footnote{\href{https://github.com/marcushogas/Cepheid-Distance-Ladder-Data}{\url{https://github.com/marcushogas/Cepheid-Distance-Ladder-Data}}}

\begin{figure}[t]
    \centering
    
    \includegraphics[width=\linewidth]{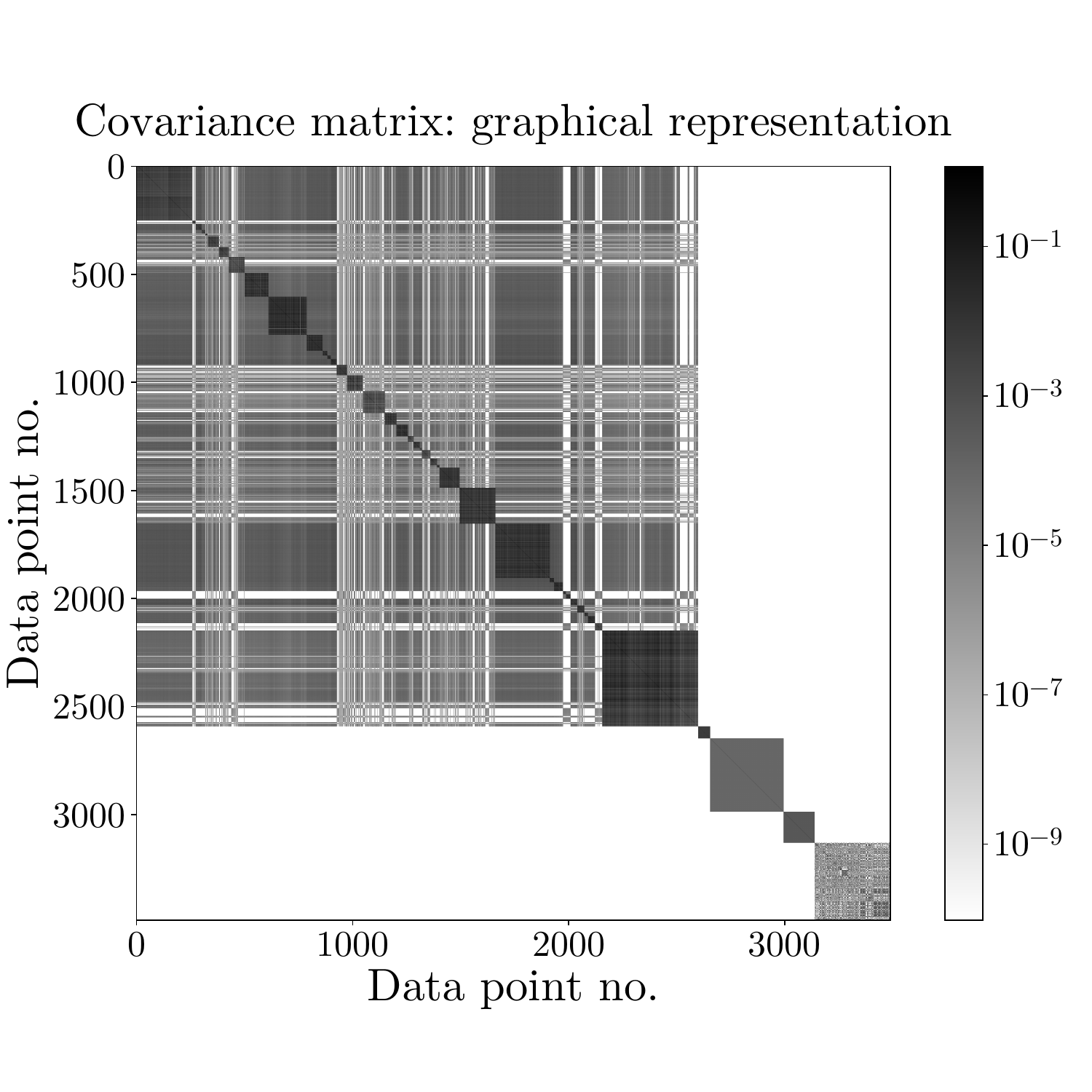}
    
    \caption{Graphical representation of the R22a covariance matrix, with the elements in the order presented in the current work.
    \label{fig:C_R22}}
\end{figure}

The expression within curly brackets $\{...\}$ in eq.~\eqref{eq:y_R22} is given in the left-hand side of eq.~\eqref{eq:mBHFSN}.
In the covariance matrix \eqref{eq:C_R22}, $\sigma_\mathrm{M101}, ... , \sigma_\mathrm{U9391}$, $\sigma_\mathrm{N4258}$, $\sigma_\mathrm{M31}$, $\sigma_\mathrm{LMC}$, $\sigma_\mathrm{SMC}$, $\sigma_\mathrm{Cal \; SN}$, and $\sigma_\mathrm{HF \; SN}$ stand for the covariance matrices pertaining to each of these data sets, including off-diagonal elements. Moreover, $Z_\mathrm{cov}$ is the covariance between Cepheid host galaxies due to a possible common systematic error in the metallicity, see \Rbase.
Fig.~\ref{fig:C_R22} displays a graphical representation of the covariance matrix.

The distance modulus to N4258 is estimated using water mega masers at the center of the galaxy \citep{Reid_2019} and the distance modulus to the LMC is estimated in \cite{Pietrzy_ski_2019} using detached eclipsing binaries.
These geometric distance estimates are implemented as external constraints in eqs.~\eqref{eq:y_R22}--\eqref{eq:L_q_R22}

In the MW, geometric distances to the Cepheids are provided via their observed parallax. 
The \Rbase calibration utilizes two sets of parallax measurements, the first due to data from the Hubble space telescope (HST) \citep{Riess:2014uga,Riess_2018} and the second from the Gaia EDR3 data set as per \cite{Riess_2021}.
The MW Cepheid photometry and periods can be found in \cite{Riess_2018,Riess:2018byc,Riess_2021}.\footnote{The HST photometry for the Cepheid SY-Aur in \cite{Riess:2014uga} as well as data for Z-Sct in \cite{Riess_2021} are not available in the literature.}
In eqs.~\eqref{eq:y_R22}--\eqref{eq:L_q_R22}, the contribution from the MW Cepheids are entered as external constraints on $\MHW$ (one for HST and one for Gaia EDR3) as well as an external constraint on $Z_W$ due to the Gaia EDR3 Cepheids. 

In \Rbase, the Cepheid metallicities, appearing in the design matrix \eqref{eq:L_q_R22}, are adopted from \cite{Romaniello:2008yh,2018A&A...619A...8G,Romaniello:2021vht,Riess:2021jrx}.
The Cepheid photometry (magnitude and color), which is needed to obtain the Wesenheit magnitudes in eq.~\eqref{eq:y_R22}, is consistently measured in the HST WFC3 photometric system to negate zero-point errors. \Rbase uses a compilation of data sets from \cite{Kato:2007zze,Soszynski:2008kd,Macri:2014xpa,Kodric:2018hpc,Riess_2019,2019wfc..rept....1R,2021ApJ...920...84L,Riess:2021jrx}. The Cepheid periods, required in the design matrix \eqref{eq:L_q_R22}, can be found in the same references.

There is one exception to the otherwise ubiquitous HST WFC3 photometry and that is a subset of the LMC Cepheids (termed ``GRND'' in eq.~\eqref{eq:L_q_R22}) and the SMC Cepheids. These two sets of Cepheids originate from the ground-based samples of \cite{Kato:2007zze,Soszynski:2008kd,Macri:2014xpa}.
To incorporate these Cepheids in the analysis, \Rbase transform the ground-based photometry into the HST system \citep{Riess:2016jrr} defining a parameter, $\Delta \mathrm{zp}$, parameterizing the difference between the HST and ground zero points. This is included as a free parameter subject to the constraint $\Delta \mathrm{zp} = 0 \pm \sigma_\mathrm{grnd}$ with $\sigma_\mathrm{grnd} = 0.10$.
This is included as an external constraint in eqs.~\eqref{eq:y_R22}--\eqref{eq:L_q_R22}.
Note that $\Delta \mathrm{zp}$ is different from $zp$ which denotes the residual parallax correction for the MW Cepheids, discussed extensively below.

In \Rbase, the SNIa redshifts, magnitudes, and covariance matrix are adopted from the Pantheon+ data sample \citep{Scolnic:2021amr}.

Since the model parameters enter linearly, the values maximizing the likelihood can be obtained analytically with the result
\begin{equation}
    q = \left( L^T C^{-1} L\right)^{-1} L^T C^{-1} y.
\end{equation}
The covariance of these parameters is contained in the matrix 
\begin{equation}
    \Sigma_q = \left( L^T C^{-1} L \right)^{-1},
\end{equation}
and the uncertainty in the model parameters can be obtained from the square root of the diagonal elements of $\Sigma_q$. The resulting $H_0$ from the \Rbase baseline calibration is,
\begin{equation}
    \label{eq:H0_R22}
    H_0 = (73.04 \pm 1.01) \, \mathrm{km/s/Mpc}. 
\end{equation}
This result is readily reproduced from the fits files published together with \Rbase or, equivalently, from the data files published together with the current paper.

\section{Recalibrating the MW Cepheids}
\label{sec:MWCephs}
The geometric distances to the MW Cepheids consist in parallax measurements, $\pi$, being related to the distance modulus as
\begin{equation}
    \label{eq:pi}
    \pi = 10^{(10 - \mu) / 5} \, \mathrm{mas}.
\end{equation}
The nonlinear relation between parallax and distance modulus prevents a linear fit without introducing simplifying assumptions. Therefore, the MW Cepheids are fitted separately in the SH0ES calibration and the resulting constraints on $\MHW$ and $Z_W$ provided as external constraints in the full distance ladder calibration, as described in Section~\ref{sec:R22}. In Section~\ref{sec:GlobalFit}, we explore the accuracy of this approach.
Before that, the next section focuses on comparing various models and statistical approaches for calibrating the MW Cepheids. Specifically, we analyze the MW Cepheids with Gaia EDR3 parallaxes, enabling a direct comparison with the results of the SH0ES team \citep{Riess_2021}. In Section~\ref{sec:AltParallaxCorr}, we explore alternatives to the conventional parallax correction.

\subsection{Gaia EDR3 Cepheids}
\label{sec:GaiaEDR3}
In this section, we re-analyze and explore different variants of the calibation of the 75 MW Cepheids featured in \cite{Riess_2021} (hereafter, \RGaia). These Cepheids exhibit HST WFC3 photometry and parallax measurements from Gaia EDR3.\footnote{The Cepheid Z-Sct was accidentally excluded from the data files and tables of \RGaia, but is retained here.}
The parallax measurements are tabulated, together with the Cepheid photometry and metallicity, in Tab.~1 of \RGaia which we adopt in the current analysis.
Of these 75 Cepheids, 66 are utilized, in conformance with \RGaia. The remaining ones either exhibit unreliable parallaxes or lie close to the boundary of Chauvenet's outlier criterion.\footnote{Including the outliers shifts the fiducial magnitude $\MHW$ by $+ 5 \, \mathrm{mmag}$ resulting in a shift of $H_0$ by $+ 0.09 \, \mathrm{km/s/Mpc}$.}

The metallicity of these Cepheids are expressed in terms of [Fe/H]. However, since we parameterize metallicity using [O/H], it is necessary to convert [Fe/H] to [O/H]. To align with \RGaia, we assume $\mathrm{[Fe/H]} = \oh$, as the conversion is not explicitly stated in that reference.

We model the magnitudes and parallaxes as
\begin{subequations}
    \begin{align}
        \label{eq:PLR2}
        m_{H,i}^W(\theta) &= \mu_i + \MHW + b_W [P]_i + Z_W [\mathrm{O/H}]_i, \\
        \label{eq:piMod}
        \pi_i(\theta) &= 10^{(10-\mu_i) / 5} - zp,
    \end{align}
\end{subequations}
cf. eq.~\eqref{eq:PLR} and eq.~\eqref{eq:pi}.
Here, and in the following, $\theta$ collectively refers to the model parameters.
As discussed in \cite{Lindegren_2021a,Lindegren_2021b}, the reported Gaia EDR3 parallaxes require corrections to account for systematic errors and achieve the desired accuracy.
These corrections, included in the tabulated values by \RGaia, are relatively well-characterized for faint sources with magnitudes $G > 13 \, \mathrm{mag}$, which are calibrated using quasars.
For objects with magnitudes $G > 10 \, \mathrm{mag}$, the correction can be extended by utilizing binary systems where the quasar-based correction applies to the fainter companion. An additional extension to brighter sources ($G < 10 \, \mathrm{mag}$) relies on binary systems where the brighter companion falls within the 6--10 mag range and the fainter one has $G > 10 \, \mathrm{mag}$.
However, as emphasized in \cite{Lindegren_2021a,Lindegren_2021b}, the accuracy of these corrections becomes uncertain in the low-magnitude regime. The bias functions ($Z_5$ and $Z_6$ in \cite{Lindegren_2021a}) were not designed to provide reliable corrections in this range, see for example Fig.~3 of \cite{Lindegren_2021a}.  
To address this limitation, and following the approach of \RGaia, we include an additional constant term, $zp$, in eq.~\eqref{eq:piMod} to account for potential systematic offsets.\footnote{$zp$ should not to be confused with $\Delta \mathrm{zp}$, which is related to the zero-point error of Cepheids with ground-based photometry.} In Section~\ref{sec:AltParallaxCorr}, we also discuss possible dependencies on the parallax correction on magnitude, color, and ecliptic latitude.

Below, we present a number of alternative calibrations, differing both with respect to the statistical methodology and the modeling.
Following \RGaia, we consider two types of models. In the first one $b_W$ and $Z_W$ are fixed and in the second $b_W$ and $Z_W$ are fitted together with the rest of the model parameters. 
Concerning the statistical methodology, we implement a number of approaches. Before we attempt to reproduce \RGaia (in the \emph{Fitting parallax only} section), we start with a comprehensive methodology, making no simplifying assumptions. Thereafter, we present a number of simplified analyses and compare them with the comprehensive one to assess their accuracy.
The results are summarized in Tab.~\ref{tab:MWcomp}.

\begin{table*}[thb]
\renewcommand*{\arraystretch}{1.6}
  \centering
  \begin{tabular}{lccccc}
    \hline\hline
    \multicolumn{6}{c}{\textbf{Model: fitting $b_W$ and $Z_W$}} \\
    Method & $M_H^W \, (\mathrm{mag})$ & $b_W \, (\mathrm{mag/dex})$ & $Z_W \, (\mathrm{mag/dex})$ & $zp \, (\mu \mathrm{as})$ & $\Delta H_0 \, (\mathrm{km/s/Mpc})$ \\
    \hline 
    \textbf{Comprehensive calibration} & $\mathbf{-5.960 \pm 0.035}$ & $\mathbf{-3.33 \pm 0.07}$ & $\mathbf{-0.18 \pm 0.13}$ & $\mathbf{-24 \pm 7}$ & $\mathbf{-0.49 (-1.47)}$ \\
    Fitting parallax only & $-5.960 \pm 0.035$ & $-3.33 \pm 0.07$ & $-0.19 \pm 0.13$ & $-24 \pm 7$ & $-0.49(-1.46)$ \\
    Linear fit & $-5.921 \pm 0.035$ & $-3.27 \pm 0.07$ & $-0.21 \pm 0.12$ & $-17 \pm 8$ & $-0.05 (-0.29)$\\
    \RGaia & $-5.915 \pm 0.030$ & $-3.28 \pm 0.06$ & $-0.20 \pm 0.13$ & $-14 \pm 6$ & $-0.06 (-0.27)$\\
    \hline\hline
    \multicolumn{6}{c}{\textbf{Model: fixing $b_W$ and $Z_W$}} \\
    Method & $M_H^W \, (\mathrm{mag})$ & $b_W \, (\mathrm{mag/dex})$ & $Z_W \, (\mathrm{mag/dex})$ & $zp \, (\mu \mathrm{as})$ & $\Delta H_0 \, (\mathrm{km/s/Mpc})$ \\
    \hline 
    \textbf{Comprehensive calibration} & $\mathbf{-5.937 \pm 0.024}$ & $-3.26^\mathrm{a}$ & $-0.17^\mathrm{a}$ &   $\mathbf{-19 \pm 6}$ & $\mathbf{-0.66 (-1.38)}$ \\
    Fitting parallax only & $-5.937 \pm 0.024$ & $-3.26^\mathrm{a}$ & $-0.17^\mathrm{a}$ & $-19 \pm 6$ & $-0.66 (-1.38)$\\
    Linear fit & $-5.920 \pm 0.025$ & $-3.26^\mathrm{a}$ & $-0.17^\mathrm{a}$ & $-16 \pm 6$ & $-0.37 (-0.83)$\\
    \RGaia & $-5.915 \pm 0.022$ & $-3.26^\mathrm{a}$ & $-0.17^\mathrm{a}$ & $-14 \pm 6$ & $0 (0)$\\
    \hline \hline
  \end{tabular}
  \caption{Inferred PLR parameters for the MW Cepheids with Gaia EDR3 parallax measurements, using different statistical methods and models. The inferred parameters are used as external constraints in the full distance ladder and we report $\Delta H_0$, the difference in the Hubble constant compared with R22a ($H_0 = 73.04 \pm 1.01$). Values within parentheses are with MW as the sole anchor galaxy. For the comprehensive calibration, we see a reduction in $H_0$ by $0.5$ -- $0.7 \, \mathrm{km/s/Mpc}$ compared with the SH0ES team's analysis. The majority of the shift in $H_0$ between our analysis and R21 is explained by the difference in the assumed prior on $zp$, with R21 assuming a normal distribution centered at zero and a $10 \, \mu \mathrm{as}$ standard deviation whereas we adopt an uninformative prior on $zp$.\\
  $^\mathrm{a}$ Fixed to the best-fit values of \cite{2019wfc..rept....1R}, following R21.}
  \label{tab:MWcomp}
\end{table*}

\subsubsection*{Comprehensive calibration}
From \RGaia, we have estimates of the parallax and Wesenheit magnitudes for 66 Cepheids in the MW. 
The corresponding uncertainties are Gaussian. Thus, the likelihood function reads
\begin{widetext}
    \begin{equation}
    \label{eq:like_full}
    \mathcal{L}(\theta) = \prod_{i=1}^{66} \frac{\exp \left[ - \frac{1}{2} \left(\frac{m^W_{H,i}(\theta) - m^{W,\mathrm{obs}}_{H,i}}{\sigma_{m,i}} \right)^2 
 - \frac{1}{2} \left(  \frac{\pi_{i}(\theta) - \pi_{\mathrm{EDR3},i}}{\sigma_{\pi,i}} \right)^2 \right]}{\sqrt{2\pi \sigma^2_{m,i} \times 2\pi \sigma^2_{\pi,i}}} .
\end{equation}
\end{widetext}
Here, $m^{W,\mathrm{obs}}_{H,i}$ and $\pi_{\mathrm{EDR3},i}$ stand for the observed magnitude and parallax, respectively, and $\sigma_{m,i}$ and $\sigma_{\pi,i}$ for their uncertainties. Note that $\sigma_{m,i}$ includes both the observational error and the intrinsic scatter of the PLR, added in quadrature. Following \RGaia, here we set the intrinsic scatter to $0.06 \, \mathrm{mag}$.\footnote{Alternatively, one can fit for the intrinsic scatter, treating it as a free model parameter. Doing so, we obtain ${\sigma_\mathrm{intr} \simeq (0.044 \pm 0.016) \, \mathrm{mag}}$, that is, slightly lower than $0.06$. However, since the intrinsic scatter is largely uncorrelated with the other PLR parameters, the effect on the calibration of the distance ladder is small, resulting in a $-0.03 \, \mathrm{km/s/Mpc}$ shift of the Hubble constant. In the rest of this section we set $\sigma_\mathrm{intr} = 0.06 \, \mathrm{mag}$, following \RGaia.}

\begin{table*}[t]
 \centering
 \renewcommand{\arraystretch}{1.2}
 \begin{tabular}{r||ccccccc}
 \hline\hline
 Parameter: & $\MHW \, (\mathrm{mag})$ & $b_W \, (\mathrm{mag/dex})$ & $Z_W \, (\mathrm{mag / dex})$ & $zp \, (\mathrm{mas})$  \\
 \hline
 Prior: & $\mathrm{U}[-7,-4]$ & $\mathrm{U}[-5,-2]$ & $\mathrm{U}[-2,2]$ & $\mathrm{U}[-0.15, 0.15]$  \\
 \hline \hline
\end{tabular}
\caption{\label{tab:prior} Uniform prior ranges for the model parameters.}
\end{table*}

To summarize, there are 70 model parameters $\theta = (\mu_i , M_H^W, b_W, Z_W, zp)$ or alternatively 68, $\theta = (\mu_i , M_H^W, zp)$, if $b_W$ and $Z_W$ are fixed. In the latter case, we follow \RGaia and fix $b_W$ and $Z_W$ to the values of \cite{2019wfc..rept....1R}, shown in Tab.~\ref{tab:MWcomp}.
The number of data points is $2 \times 66 = 132$. With the likelihood set up in eq.~\eqref{eq:like_full} we constrain the model parameters using a Bayesian approach, imposing uninformative priors on the model parameters (i.e., uniform and with a wide range), see Tab.~\ref{tab:prior}. The posterior distribution of the model parameters is inferred using the Markov Chain Monte Carlo (MCMC) method of \cite{GoodmanWeare}, as implemented in the \texttt{emcee} Python package of \cite{foreman2013emcee}.
For convergence of the chains, we demand that the length of the chains exceed 100 times the estimated autocorrelation time ($\tau$). Additionally, the chains are inspected visually to confirm the convergence.
We designate the initial $2 \tau$ samples as burn-in and exclude them from the final analysis. The number of walkers is twice the number of model parameters. These procedures are consistently applied to all subsequent MCMC iterations in this study.

\begin{figure}[t]
    \centering
    
    \includegraphics[width=\linewidth]{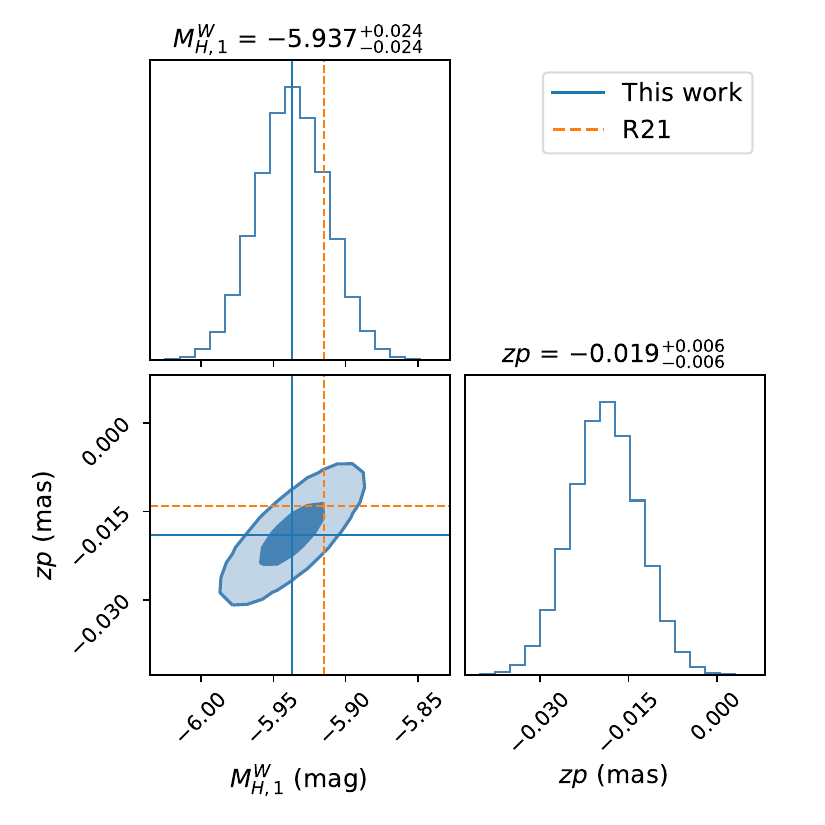}
    
    \caption{Marginalized posterior distribution for $\MHW$ and $zp$ when fitting the Gaia EDR3 MW Cepheids with a model fixing $b_W$ and $Z_W$ to the values of \cite{2019wfc..rept....1R}. $39 \, \%$ and $86 \, \%$ credence regions, corresponding to $1 \, \sigma$ and $2 \, \sigma$. There is a clear degeneracy between the fiducial magnitude $\MHW$ and the residual parallax offset $zp$. The orange, dashed line indicates the R21 results.
    \label{fig:GaiaEDR3_corner}}
\end{figure}

The results are presented in Tab.~\ref{tab:MWcomp}.
For the model with $b_W$ and $Z_W$ fixed, we obtain a fiducial magnitude, $\MHW$, that is $1.0 \, \sigma$ lower than that in \RGaia. For the model with $b_W$ and $Z_W$ fitted, the corresponding number is $1.5 \, \sigma$ lower.
When using this as an external constraint in the full distance ladder, it leads to an $H_0$ which is $\simeq 0.6 \, \mathrm{km/s/Mpc}$ lower than the baseline analysis of \Rbase.

The marginalized posterior distribution for $\MHW$ and $zp$ is shown in Fig.~\ref{fig:GaiaEDR3_corner} for a model with $b_W$ and $Z_W$ fixed, revealing that $\MHW$ and $zp$ have a positive correlation. Thus, we also obtain a residual parallax offset which is significantly lower than that of \RGaia. With $b_W$ and $Z_W$ fixed, we obtain a $> 3 \, \sigma$ evidence for $zp$ being negative whereas the evidence is $\simeq 2 \, \sigma$ in \RGaia. 

When $b_W$ and $Z_W$ are fitted, the $\chi^2$-value for the best-fit parameters is $\chi^2_\mathrm{min} = 64.7$ for $132 - 70 = 62$ degrees of freedom, corresponding to a reduced $\chi^2$-value of $\chi^2_\mathrm{dof} = 1.04$.
When $b_W$ and $Z_W$ are fixed, the corresponding numbers are $\chi^2_\mathrm{min} = 65.6$ for $132 - 68 = 64$ degrees of freedom, that is, $\chi^2_\mathrm{dof} = 1.03$.
The results indicate reasonable fits in both cases.

In the next section, where we follow a statistical method similar to \RGaia, we explain the difference between our results and \RGaia.

As a check, we also performed a frequentist analysis, with results being consistent with the Bayesian approach presented here.

\subsubsection*{Fitting parallax only}
A simplified approximation of the likelihood \eqref{eq:like_full} can be obtained by expanding $m_{H,i}^W$ to first order around $m_{H,i}^{W,\mathrm{obs}}$, then marginalizing over the magnitude. A full derivation is presented in Appendix~\ref{sec:like_parallax}. The result is
\begin{equation}
    \label{eq:like_parallax}
    \mathcal{L}(\theta) = \prod_{i=1}^{66} \frac{\exp \left[ - \frac{1}{2} \left(  \frac{\pi_{i}^\mathrm{phot}(\theta) - \pi_{\mathrm{EDR3},i}}{\sigma_{\mathrm{tot},i}} \right)^2 \right]}{\sqrt{ 2\pi \sigma^2_{\mathrm{tot},i}}},
\end{equation}
where 
\begin{subequations}
\label{eq:parallax_model}
    \begin{align}
        \pi_i^\mathrm{phot}(\theta) &= 10^{(10 - \mu_i^\mathrm{phot}(\theta)) / 5} - zp,\\
        \mu_i^\mathrm{phot}(\theta) &= m_{H,i}^{W,\mathrm{obs}} - \MHW - b_W \p_i - Z_W \oh_i,
    \end{align}
\end{subequations}
and the total error $\sigma_{\mathrm{tot},i}$ includes the error in the parallax as well as an error due to the uncertainty in the magnitude, that is,
\begin{equation}
    \label{eq:sigtot}
    \sigma_{\mathrm{tot},i}^2 = \sigma_{\pi, i}^2  + \left( 10^{-\mu_i^\mathrm{phot}(\theta)/5} \, 20 \ln 10 \, \sigma_{m,i} \right)^2.
\end{equation}
Recall that $\sigma_{m,i}$ includes both the observational uncertainty as well as intrinsic scatter, added in quadrature.

Assuming that our model provides a good fit to data, we expect that fitting the parallax only, with the likelihood \eqref{eq:like_parallax}, should yield a result approximating the comprehensive calibration.

Since we are fitting the parallax, the model is still nonlinear and we use the same MCMC method as in the comprehensive calibration to infer the posterior distribution, only now with 66 fewer model parameters---a huge gain in computational cost.
The result is presented in Tab.~\ref{tab:MWcomp} and is virtually identical with that of the comprehensive calibration.
Thus, we have shown that the method of fitting the parallaxes yields the same precision and accuracy as the comprehensive calibration, so it can readily be used.

\RGaia also infer the model parameters by fitting the parallax but, as shown in Tab.~\ref{tab:MWcomp}, there is a significant difference from the results presented here. There are two reasons for this:
\begin{itemize}
    \item First, and most importantly, we impose an uninformative prior on $zp$ whereas \RGaia use a Gaussian prior centered at $0 \, \mu \mathrm{as}$ with a standard deviation of $10 \, \mu \mathrm{as}$. 
    \RGaia justify their choice based on the statement in \cite{Lindegren_2021a}: ``It is difficult to quantify uncertainties in $Z_5$ and $Z_6$. In the region of the parameter space that is well populated by the quasars (essentially $G \gtrsim 16$ and $1.4 \lesssim \nu_\mathrm{eff} \lesssim 1.7 \, \mu \mathrm{m}^{-1}$), they may be as small as a few $\mu$as, but beyond that region, uncertainties are bound to be greater because of the indirect methods we used.'' Given that the residual parallax offset is expected to be only a few microarcseconds in this well-calibrated regime, \RGaia extend this estimate to the high-brightness range of Cepheids by assuming a proportional increase in uncertainty, scaling it up by a factor of a few.
    
    \item Second, the normalization $( 2 \pi \sigma_{\mathrm{tot},i}^2 )^{-1/2}$ in the likelihood~\eqref{eq:like_parallax} is absent in \RGaia. However, this must be included due to the dependence of $\sigma_{\mathrm{tot},i}$ on the model parameters, cf. eq.~\eqref{eq:sigtot}. That is, $\sigma_{\mathrm{tot},i}$ is not constant.
\end{itemize}
The two differences result in a total $\simeq - 0.6 \, \mathrm{km/s/Mpc}$ shift in $H_0$, as seen in Tab.~\ref{tab:MWcomp}, whereof the choice of $zp$ prior contributes with $\simeq 75 \, \%$ of the shift and the normalization term with the remaining $\simeq 25 \, \%$.
Setting a prior of $zp = \mathcal{N}(0, 10) \, \mu \mathrm{as}$ and omitting the normalization of the likelihood, we reproduce the \RGaia results.

The choice of prior on $zp$ significantly influences the calibration of the distance ladder, given the reported uncertainty in $H_0$ of $0.9 \, \mathrm{km/s/Mpc}$.
In \cite{Lindegren_2021a,Lindegren_2021b}, which provide the original parallax corrections, the authors explicitly acknowledge the uncertainty in these corrections for the low-magnitude range where Cepheids are located. For instance, they note: ``\emph{While it is easy enough to demonstrate that the EDR3 parallaxes
contain significant systematics, it is extremely difficult to obtain accurate estimates of the bias beyond the limited region of parameter space that is well populated by the quasars.}'' \citep[p.~23]{Lindegren_2021a}.
Recognizing these challenges, the authors encourage the use of external calibrations for the residual parallax offset. 

Several external calibrations of $zp$ are available in the literature. \cite{Stassun:2021} suggests a positive value for $zp$. However, the large uncertainty in the latter makes it consistent with the value $zp \simeq -20 \, \mu\mathrm{as}$ found in \cite{Bhardwaj:2021,Fabricius:2021}.
\cite{Huang:2021,Zinn:2021} also favor a negative residual offset but are somewhat inconsistent with \cite{Bhardwaj:2021}.
Overall, most external calibrations suggest a negative parallax offset somewhere between $0 \, \mu \mathrm{as}$ and $-20 \, \mu \mathrm{as}$. Therefore, considering the uncertainty in the parallax correction, as highlighted in \cite{Lindegren_2021a,Lindegren_2021b}, and the minor discrepancies among these external calibrations, we adopt a conservative approach rather than imposing the prior advocated by \RGaia. Specifically, we impose an uninformative prior on $zp$, allowing the distance ladder calibration to determine its value. This method also enables us to provide an independent external calibration of the residual parallax offset within this magnitude range.

\subsubsection*{Linear fit}
The uncertainties in observed parallaxes are Gaussian. Due to the nonlinear relation between parallax and distance, eq.~\eqref{eq:pi}, the corresponding uncertainties in the distance moduli are non-Gaussian.
In \Rbase, it is thus argued that to avoid a bias, known as the Lutz--Kelker bias, one should fit the parallax rather than the distances.
Nevertheless, here we explore the consequences of fitting the distances, which has the advantage that it can be done linearly.

Solving eq.~\eqref{eq:piMod} for $\mu$, we obtain,
\begin{equation}
    \mu = 10 - 5 \log \pi -5 \log \left( 1 + \frac{zp}{\pi} \right).
\end{equation}
We expect $zp \sim 0.01 \, \mathrm{mas}$. Thus, $zp / \pi \sim 0.01$. Making a Taylor expansion around $zp / \pi = 0$ to second order, we get,
\begin{equation}
    \mu \simeq 10 - 5 \log \pi - \frac{5}{\ln 10} \frac{zp}{\pi} + \frac{5}{2 \ln 10} \left( \frac{zp}{\pi} \right)^2 .
\end{equation}
Plugging into eq.~\eqref{eq:PLR2}, we get
\begin{align}
    \label{eq:fit_lin}
    m_{H,i}^W - 10 + 5 \log \pi_{\mathrm{EDR3},i} - \frac{5}{2 \ln 10} \left( \frac{zp_0}{\pi_{\mathrm{EDR3},i}} \right)^2 = \nonumber \\
    \MHW + b_W \p_i + Z_W \oh_i +zp \left( - \frac{5}{\pi_{\mathrm{EDR3},i} \ln 10} \right) .
\end{align}
On the left-hand side, we have the data and $zp_0$ is an initial guess for the residual parallax correction, $zp_0$ is subsequently updated iteratively with the inferred value of $zp$ until convergence is reached, that is, when $zp_0$ equals the inferred value to some precision, here $\pm 0.5 \, \mu \mathrm{as}$.

On the right-hand side of eq.~\eqref{eq:fit_lin} we have the model prediction with the model parameters $(\MHW, b_W, Z_W, zp)$.\footnote{If $b_W$ and $Z_W$ are fixed, these terms are moved to the left-hand side of eq.~\eqref{eq:fit_lin}.}
Note that all model parameters appear linearly. Thus, just as in Section~\ref{sec:R22}, the model parameters can be inferred linearly. The results are presented in Tab.~\ref{tab:MWcomp} from which we note that the method of fitting distances linearly yields significantly different results for the inferred model parameters, compared with the full nonlinear fit.

\subsubsection*{Comparison of methods: recommendation}
When calibrating the PLR using MW Cepheids, the comprehensive calibration does not make simplifying assumptions.
Fitting the parallax is also accurate while having the advantage of being significantly faster, but still requires MCMC methods.
The linear method is the fastest, but yields biased results. 
Therefore, we recommend to fit the MW parallaxes.
Our analysis results in a PLR zero-point ($\MHW$) which is $1.0$ -- $1.5 \, \sigma$ lower than that of \RGaia resulting in a $0.5$ -- $0.7 \, \mathrm{km/s/Mpc}$ decrease in the Hubble constant, or a $1.4$ -- $1.5 \, \mathrm{km/s/Mpc}$ decrease if the MW is used as the sole anchor.

\subsection{Alternative parallax corrections}
\label{sec:AltParallaxCorr}
As explained in \cite{Lindegren_2021a}, to be accurate the reported Gaia EDR3 parallaxes are corrected for their magnitude, color, and ecliptic latitude. These corrections are all included in the tabulated parallax values of \RGaia that we are using here.
Following \RGaia, in eq.~\eqref{eq:piMod} we added a constant residual parallax offset to account for the uncertainty of the parallax corrections of \cite{Lindegren_2021a}.
In this section, we consider alternative residual parallax corrections, depending on the magnitude, color, and ecliptic latitude. The respective models read,
\begin{subequations}
    \begin{align}
        \label{eq:zp_alt2}
        \pi &= 10^{(10 - \mu)/5} - zp_1 -zp_2 \, m_H, \\
        \label{eq:zp_alt3}
        \pi &= 10^{(10 - \mu)/5} - zp_1 -zp_2 \, (m_V - m_I), \\
        \label{eq:zp_alt4}
        \pi &= 10^{(10 - \mu)/5} - zp_1 -zp_2 \, \sin \beta,
    \end{align}
\end{subequations}
where $\beta$ is the ecliptic latitude. 

\begin{figure*}[t]
    \centering
    
    \includegraphics[width=0.49\linewidth]{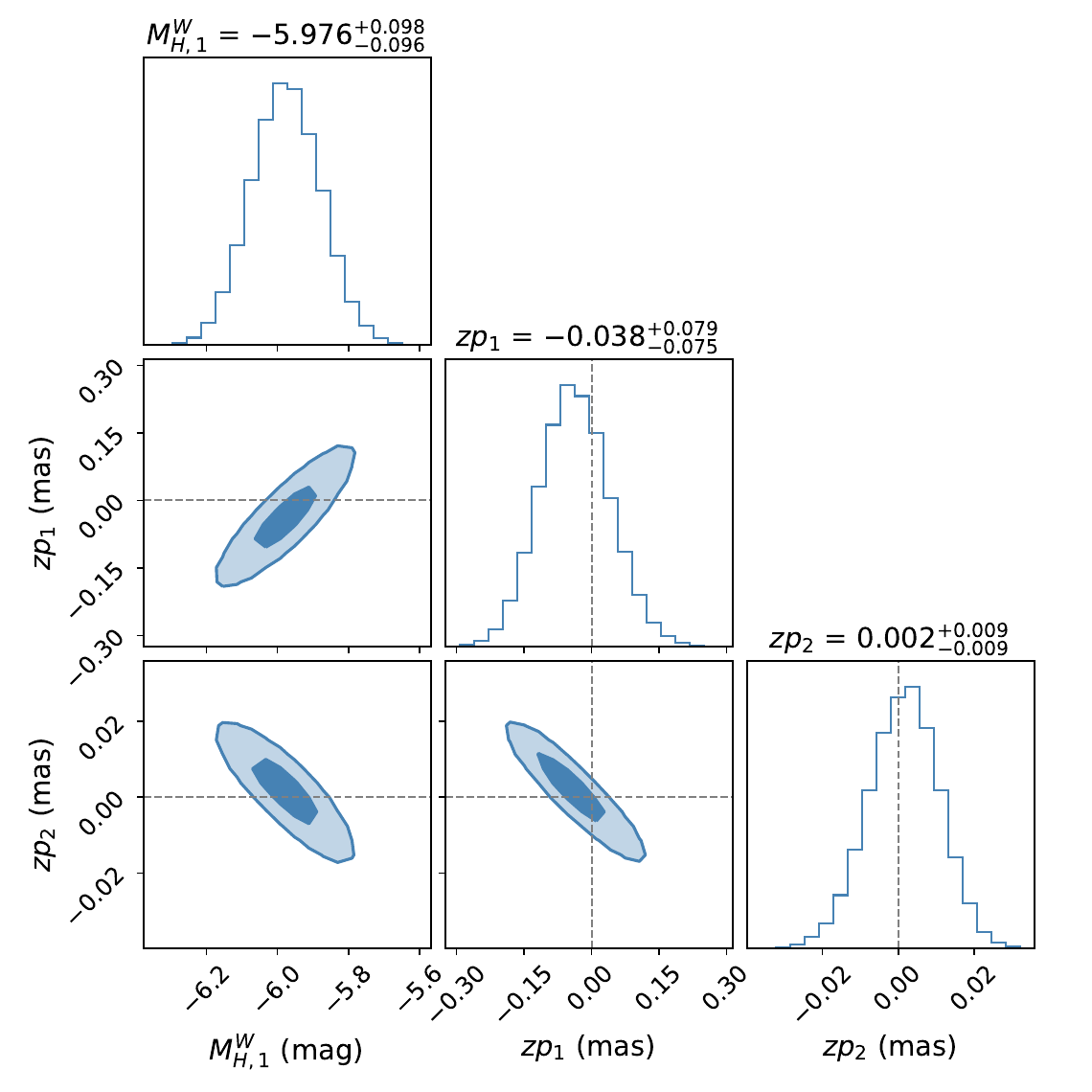}
    \includegraphics[width=0.49\linewidth]{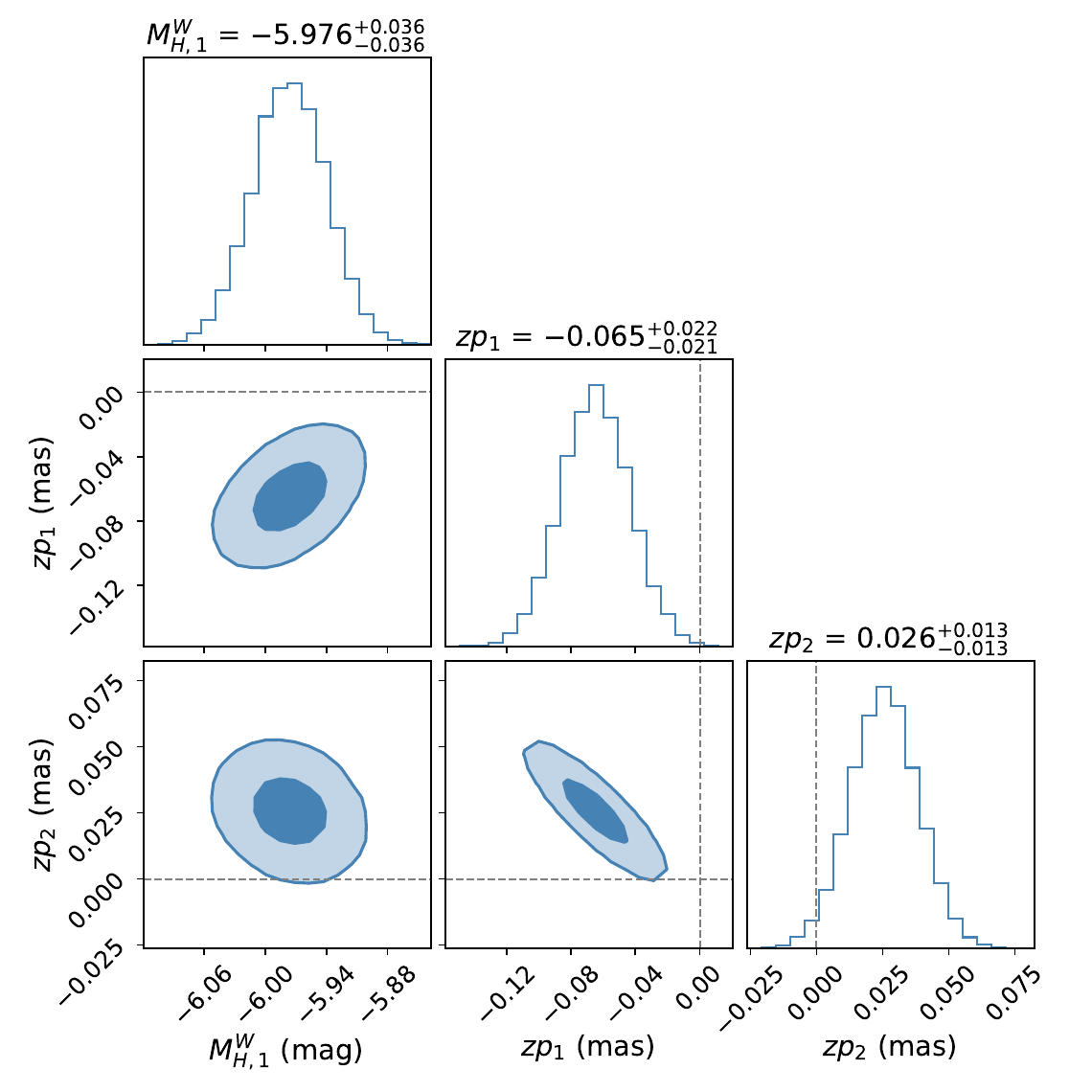}
    
    \caption{Alternative parallax corrections for the 66 MW Cepheids with Gaia EDR3 parallax measurements. 2D marginalized confidence contours ($39 \, \%$ and $86 \, \%$ credence regions, corresponding to $1 \, \sigma$ and $2 \, \sigma$, respectively). The gray, dashed lines indicate a null result for the residual parallax offsets. \emph{Left:} Magnitude-based correction, eq.~\eqref{eq:zp_alt2}. Consistent with no residual parallax offset. \emph{Right:} Color-based correction, eq.~\eqref{eq:zp_alt3}. Here, there is a preference for non-zero residual parallax offsets.
    \label{fig:MWCeph_alt}}
\end{figure*}

We fit the 66 MW Cepheids with Gaia EDR3 parallax measurements, as described above. Here, we are interested in the possible evidence for alternative parallax corrections from the MW Cepheids alone. Thus, we fit $b_W$ and $Z_W$ together with the other parameters rather than fixing them, so $\theta = (\MHW,b_W,Z_W,zp_1,zp_2)$. 
The resulting marginalized posterior distributions are shown in Fig.~\ref{fig:MWCeph_alt}.

\begin{figure}[t]
    \centering
    
    \includegraphics[width=\linewidth]{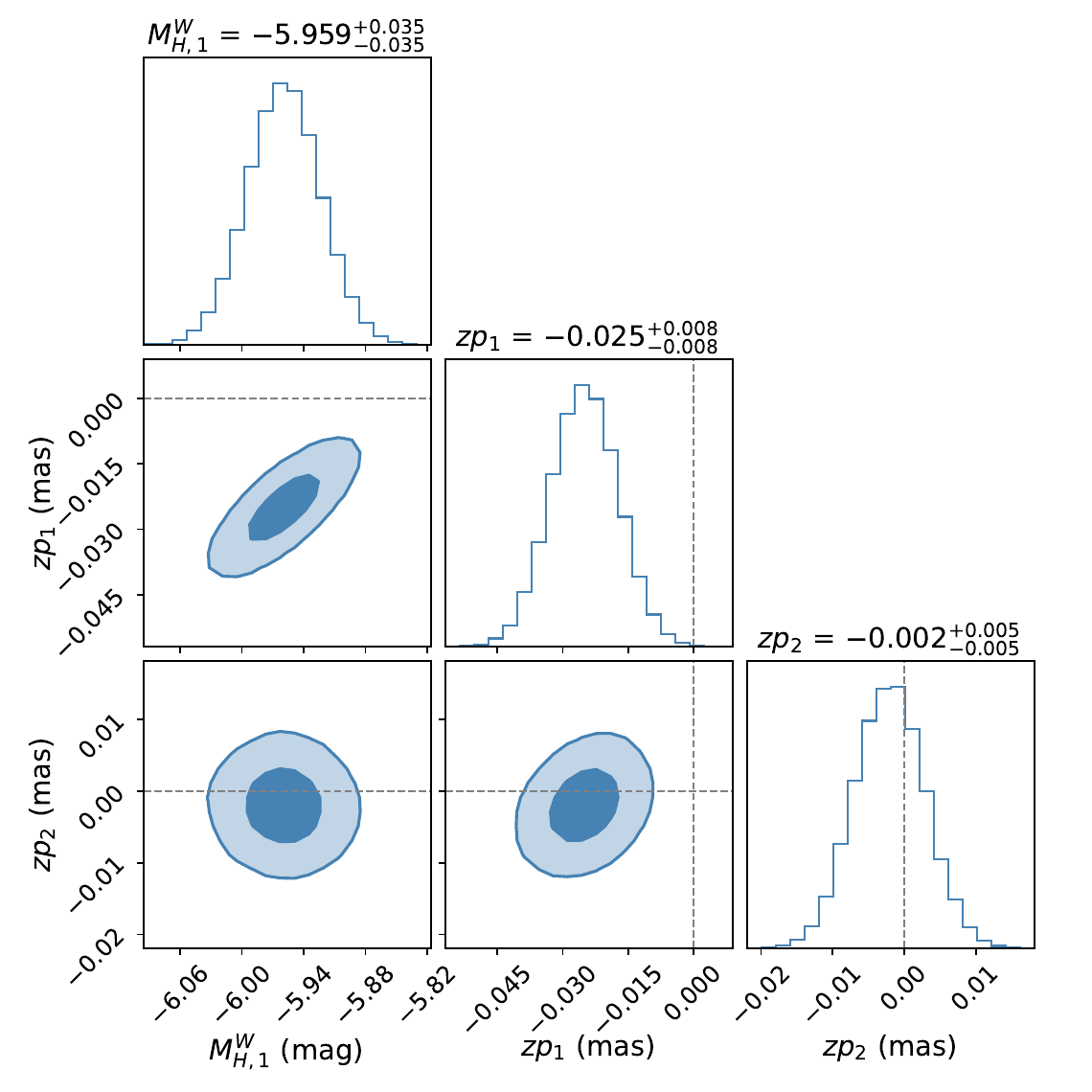}
    
    \caption{Same as Fig.~\ref{fig:MWCeph_alt} but for the ecliptic latitude-based correction, eq.~\eqref{eq:zp_alt4}. Consistent with no residual parallax offset.
    \label{fig:MWCeph_alt2}}
\end{figure}

For the first, magnitude-dependent, correction \eqref{eq:zp_alt2}, $zp_2$ is consistent with zero, that is, such a correction does not seem to be justified. 
The most notable difference is an almost threefold increase in the uncertainty in the inferred value of $\MHW$, rendering the MW contribution to the distance ladder calibration of $H_0$ subdominant compared with the remaining anchor galaxies. 

For the second correction \eqref{eq:zp_alt3}, $zp_2$ is non-zero with a $2 \, \sigma$ significance. Thus, with this model one achieves an overall better fit of the MW Cepheids. 
However, comparing the Bayesian information criterion (BIC) of this model with that of a constant parallax correction, there is no strong evidence neither for nor against this additional color-dependent correction.

With a color-dependent correction, $\MHW$ is shifted $\simeq - 0.02 \, \mathrm{mag}$ and its uncertainty is almost unchanged. Updating the external constraint $M^W_{H,1,\mathrm{Gaia}}$ in eqs.~\eqref{eq:y_R22}--\eqref{eq:L_q_R22} with this value leads to a total shift in $H_0$ by $-0.61 \, \mathrm{km/s/Mpc}$ compared with the SH0ES team, when calibrating the full distance ladder.
We obtained the constraint on $\MHW$ by fitting $b_W$ and $Z_W$ together with $\MHW$, $zp_1$, and $zp_2$. So, the present shift in the Hubble constant should be compared with that of the first line of Tab.~\ref{tab:MWcomp}, that is, $-0.49 \, \mathrm{km/s/Mpc}$.
We conclude that a color-dependent parallax correction leads to an additional $\simeq -0.1 \, \mathrm{km/s/Mpc}$ shift of $H_0$, that is, a minor effect.

For the third correction eq.~\eqref{eq:zp_alt4} based on ecliptic latitude, $zp_2$ is consistent with zero. Thus, such a correction appears superfluous. 
Moreover, the best-fit value of the magnitude aligns almost perfectly with the corresponding fit with a constant parallax offset, see Tab.~\ref{tab:MWcomp}. Hence such a correction does not have any effect on the $H_0$ estimate.

\subsection{Global fit}
\label{sec:GlobalFit}
In this section, we present a comprehensive global fit of the Cepheid-based distance ladder, incorporating Milky Way Cepheids directly into the fitting process. This approach contrasts with the SH0ES teams' method, treating MW Cepheids as external constraints. By fitting all components of the distance ladder simultaneously, we aim to evaluate the impact of integrating the MW Cepheids via external constraints in the global analysis and to assess whether fixing or fitting the parameters $b_W$ and $Z_W$ provides better accuracy.

In this section, we use the most updated SH0ES team data, including the following MW Cepheid data sets:
\begin{itemize}
    \item \textbf{Gaia EDR3 MW Cepheids:} 66 Cepheids with Gaia parallax measurements, discussed in detail in Section~\ref{sec:GaiaEDR3}.
    \item \textbf{HST MW Cepheids:} 8 Cepheids with HST parallaxes measurements sourcing the external constraint $M^W_{H,1,\mathrm{HST}}$ in the \Rbase calibration, cf. eqs.~\eqref{eq:y_R22}--\eqref{eq:L_q_R22}. Seven parallaxes are adopted from \cite{Riess_2018} and one from \cite{Riess:2014uga}.\footnote{The HST photometry for SY-Aur is unfortunately unavailable and thus not included in the present analysis.} 
    \item \textbf{Cluster MW Cepheids:} 17 Cepheids in open clusters \citep{Riess:2022mme}. The parallax is estimated from the stars in the open cluster using the Gaia EDR3 data set. We adopt the parallax values and HST photometry presented in Tab.~1 and Tab.~2 of \cite{Riess:2022mme}.
\end{itemize}
The residual parallax offset of the Gaia EDR3 MW Cepheids ($zp_\mathrm{Gaia}$) was discussed above where we argued for a conservative approach by setting an uninformative prior on it. 
Regarding the residual parallax offset of the cluster MW Cepheids ($zp_\mathrm{cluster}$), it is independent of $zp_\mathrm{Gaia}$.
The parallaxes of the cluster Cepheids are determined by their associated cluster stars, which span the magnitude range $8 \, \mathrm{mag} \lesssim G \lesssim 17 \, \mathrm{mag}$.
Notably, some of these stars fall within the $G \gtrsim 16 \, \mathrm{mag}$ regime, where the parallax offset is expected to be smaller than in the lower-magnitude regime.
As noted in \cite{Lindegren_2021a}, the uncertainty in parallax corrections in the $G \gtrsim 16 \, \mathrm{mag}$ range ``may be as small as a few $\mu$as''.

This raises the question of whether a prior should be imposed on $zp_\mathrm{cluster}$. However, several factors temper our confidence in adopting such a prior. First, \cite{Lindegren_2021a} does not specify a precise uncertainty range (beyond ``a few $\mu$as''), nor does the phrasing (``may be'') suggest sufficient confidence to justify an informative prior.
Second, the $G \gtrsim 16$ regime, where the parallax corrections are likely associated with smaller uncertainties, only partially overlaps with the magnitude range of the cluster stars, complicating the application of such a prior. 
Finally, imposing a prior of $zp_\mathrm{cluster} = 0 \pm 10 \, \mu \mathrm{as}$ does not significantly alter $H_0$, resulting in a shift of $+0.1 \, \mathrm{km/s/Mpc}$.

For these reasons, we adopt the same noninformative prior on $zp_\mathrm{cluster}$ as we do for $zp_\mathrm{Gaia}$ in the subsequent analysis, allowing for an independent calibration of the parallax offset in the $8 \, \mathrm{mag} \lesssim G \lesssim 17 \, \mathrm{mag}$ regime.

From Tab.~1 of \cite{Bhardwaj:2023mau} we adopt updated metallicity values for 42 of the MW Cepheids and propagate their uncertainty in metallicity into the total parallax uncertainty.
Following \cite{Riess:2022mme}, the MW metallicity estimates are converted from [Fe/H] to [O/H] using the relation $\oh = [\mathrm{Fe/H}] + 0.06$. 

The remaining distance ladder data is taken from the fits files published together with \Rbase\footnote{That is, as described in Section~\ref{sec:R22}. The corresponding data files are available at: \href{https://github.com/marcushogas/Cepheid-Distance-Ladder-Data}{\url{https://github.com/marcushogas/Cepheid-Distance-Ladder-Data}}.} with the following adaptation and updates:
\begin{itemize}
    \item \textbf{MW constraints:} Since the MW Cepheids are included directly in this global fit, their contributions as external constraints ($M_{H,\mathrm{Gaia}}^W$, $M_{H,\mathrm{HST}}^W$, $Z_W$) are removed from eqs.~\eqref{eq:y_R22}--\eqref{eq:L_q_R22}.
    \item \textbf{SMC photometry and distance:} We replace the \Rbase SMC photometry with the updated HST WFC3 photometry contained in Tab.~2 of \cite{Breuval:2024lsv}. 
    The SMC metallicities have also been updated by Romaniello et al. (in prep.) but is currently unavailable. Instead, we use the reported mean value $\oh = -0.585 \, \mathrm{dex}$ in \cite{Breuval:2024lsv}.    
    The diagonal element in the SMC covariance matrix, corresponding to Cepheid number $i$, is given by $\sigma_{m,i}^2 + \sigma_\mathrm{intr}^2 + \sigma_{\mathrm{geom}}^2$, where $\sigma_{m,i}$ is the photometric uncertainty, $\sigma_\mathrm{intr}$ is the intrinsic scatter of the PLR, and $\sigma_\mathrm{geom} = 0.05 \, \mathrm{mag}$ is the uncertainty due to a geometric correction \citep{Breuval:2024lsv}.
    The off-diagonal elements of the SMC covariance matrix are set to $10^{-4}$, accounting for systematic uncertainty in the metallicity abundance scale (following eq.~(9) of \Rbase).
    Additionally, the SMC’s geometric distance is applied with the relative distance between SMC and LMC being well determined in \cite{Graczyk_2020}:
    \begin{equation}
        \Delta \mu_\mathrm{SMC/LMC} = 0.500 \pm 0.017 \, \mathrm{mag}.
    \end{equation}
    Thus, the SMC is used as an anchor galaxy, which was not the case in \Rbase.\footnote{The SMC distance was cited in \Rbase by never used in the actual fit.}
    \item \textbf{SNIa covariance update:} The SNIa covariance matrix is updated per \cite{Murakami:2023xuy}, reducing scatter in the supernova magnitudes, leading to a $14 \, \%$ reduction in the uncertainty of $H_0$. 
\end{itemize}
Following \Rbase, we set the intrinsic PLR scatter $\sigma_\mathrm{intr} = 0.07 \, \mathrm{mag}$ for this global fit.
The likelihood has four contributions, three from the MW Cepheids and one from the remaining distance ladder:
\begin{enumerate}
    \item \textbf{Gaia EDR3 MW Cepheids:} The likelihood is given by eqs.~\eqref{eq:like_parallax}--\eqref{eq:parallax_model} with a residual parallax offset that we refer to here as $zp_\mathrm{Gaia}$.
    \item \textbf{Cluster MW Cepheids:} The likelihood follows the same form as Gaia EDR3 Cepheids, but with an independent residual parallax offset, $zp_\mathrm{cluster}$.
    \item \textbf{HST MW Cepheids:} The likelihood follows eqs.~\eqref{eq:like_parallax}--\eqref{eq:parallax_model}, with the residual parallax offset set to zero, since this method of estimating the parallax is not expected to exhibit the same type of offset as those provided by the Gaia satellite.
    \item \textbf{Remaining distance ladder:} The likelihood for the remaining distance ladder is obtained by $\ln \mathcal{L}(\theta) = -0.5 \, \chi^2(\theta)$ with
    \begin{equation}
        \chi^2(\theta) = \left[ y - Lq(\theta) \right]^T C^{-1} \left[ y-Lq(\theta) \right],
    \end{equation}
    cf. eq.~\eqref{eq:chi2_R22}.\footnote{Here, we have omitted the normalization of $\mathcal{L}$. This can be done without loss of generality since here it is a constant that does not depend on the model parameters.}
\end{enumerate}
In Appendix~\ref{sec:GlobalFitLikelihood}, we present a comprehensive account of the full likelihood.

In summary, this analysis exhibits 49 model parameters: 37 distance moduli to the SNIa host galaxies, the distances to the anchor galaxies $\mu_\mathrm{N4258}$, $\mu_\mathrm{LMC}$, $\Delta \mu_\mathrm{SMC/LMC}$, the distance to M31, $\mu_\mathrm{M31}$, the PLR intercept and slopes $\MHW$, $b_W$, and $Z_W$, the residual parallax offset for the Gaia EDR3 Cepheids and the MW cluster Cepheids $zp_\mathrm{Gaia}$ and $zp_\mathrm{cluster}$, the difference between the HST and ground zero points $\Delta \mathrm{zp}$, the fiducial SNIa absolute magnitude $M_B$, and finally the Hubble constant, via the parameter $5 \log H_0$. 
The number of data points is 3523. 
Since the MW Cepheids are included in a global calibration, it is not possible to solve linearly, instead we use an MCMC method making a full nonlinear fit for all model parameters (as described above).

\begin{figure*}[t]
    \centering
    
    \includegraphics[width=\linewidth]{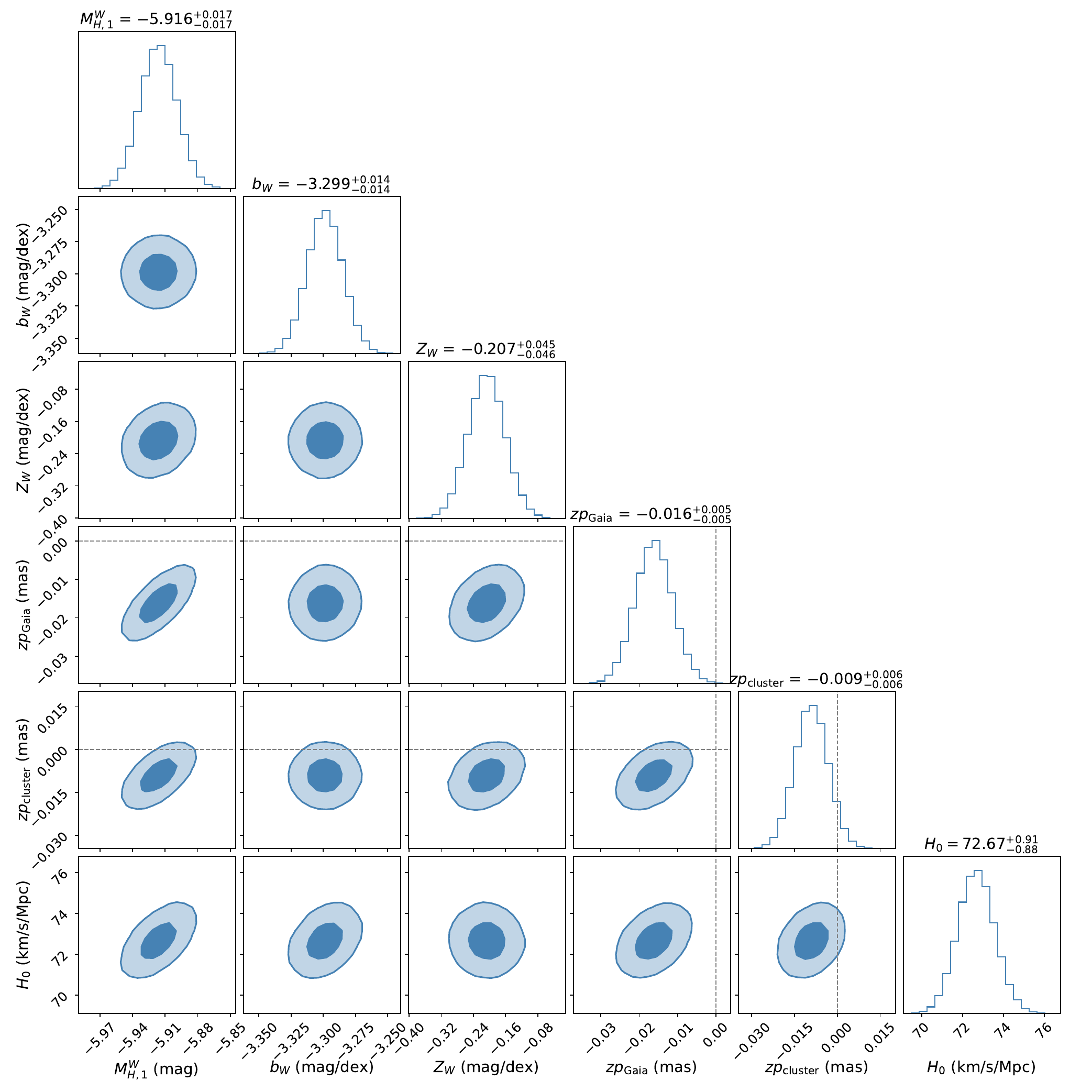}
    
    \caption{Marginalized posterior distributions for a selection of model parameters, derived from a comprehensive calibration of the latest Cepheid-based distance ladder data. The analysis includes a simultaneous fit of Milky Way Cepheids alongside data from the rest of the distance ladder. Our analysis exhibits a $- 0.50 \, \mathrm{km/s/Mpc}$ shift in the Hubble constant compared with the corresponding SH0ES team result \citep{Breuval:2024lsv}. Shaded regions indicate the $39 \, \%$ and $86 \, \%$ credible intervals, corresponding to the $1 \, \sigma$ and $2 \, \sigma$ confidence levels, respectively. The gray, dashed lines indicate zero residual parallax offset. There is a $\simeq 3 \, \sigma$ significance for a negative $zp_\mathrm{Gaia}$ while $zp_\mathrm{cluster}$ is consistent with zero, within $2 \, \sigma$. 
    \label{fig:GlobalFit}}
\end{figure*}

The 2D marginalized posterior distribution for a selection of model parameters is shown in Fig.~\ref{fig:GlobalFit}. 
The inferred value for the Hubble constant is
\begin{equation}
    \label{eq:H0_FullGlobal}
    H_0 = (72.67 \pm 0.90) \, \mathrm{km/s/Mpc} .
\end{equation}
This should be compared with $H_0 = (73.17 \pm 0.86) \, \mathrm{km/s/Mpc}$ which is reported in \cite{Breuval:2024lsv} and includes the same updates as in the present fit. 
We attribute the $- 0.50 \, \mathrm{km/s/Mpc}$ downwards shift to the differing MW analysis, as discussed extensively in the sections above. 
This reduction is slightly less than the $-0.66 \, \mathrm{km/s/Mpc}$ shift reported in Tab.~\ref{tab:MWcomp} for the model with $b_W$ and $Z_W$ fixed which, as we argue below, is the most accurate model. The lesser shift here is due to the fact that we have used the most updated MW data in this section, including MW cluster Cepheids and updated metallicities.

The residual parallax offsets are 
\begin{equation}
    zp_\mathrm{Gaia} = (-16 \pm 5) \, \mu \mathrm{as}, \quad zp_\mathrm{cluster} = (-9 \pm 6) \, \mu \mathrm{as}.
\end{equation}
The former is significant at $\simeq 3 \, \sigma$, while the latter is consistent with zero.
The latter is expected since the cluster stars used to estimate the Gaia EDR3 parallax exhibit magnitudes ($G > 13$) in the range where the Gaia calibration is most comprehensive. 

Now, we compare the global fit with simplified approaches where MW Cepheids are calibrated separately:
\begin{itemize}
    \item \textbf{Fixed $b_W,Z_W$:} When the parameters $b_W$ and $Z_W$ are fixed during the MW Cepheid calibration, the resulting $\MHW$ is used as an external constraint in the full distance ladder. This approach yields $H_0 = (72.73 \pm 0.89) \, \mathrm{km/s/Mpc}$ which deviates by only $0.06 \, \mathrm{km/s/Mpc}$ from the global fit. In this case, the MW Cepheids are fitted with fixed $b_W$ and $Z_W$, and the $\MHW$ constraint is used in the full distance ladder analysis.
    \item \textbf{Fitted $b_W,Z_W$:} In this approach, both $b_W$ and $Z_W$ are treated as free parameters during the MW Cepheid calibration. When these parameters are included in the fit for the full distance ladder, we obtain $H_0 = (72.35 \pm 0.88) \, \mathrm{km/s/Mpc}$, a larger deviation of $-0.32 \, \mathrm{km/s/Mpc}$ compared to the global fit.
\end{itemize}
To balance computational efficiency with accuracy, we therefore recommend calibrating MW Cepheids separately with $b_W$ and $Z_W$ fixed, and using the resulting $\MHW$ as an external constraint in the broader distance ladder calibration, which can be done linearly. This method produces results comparable to a full global fit but at a reduced computational cost.

\subsection{Summary of MW results}
Our renewed MW analysis exhibits a $- 0.5 \, \mathrm{km/s/Mpc}$ shift in $H_0$ compared with the SH0ES team. The reduction is mostly explained by our choice of an uninformative prior on the residual parallax offset whereas \RGaia use a prior centered at zero with an uncertainty of $10 \, \mu \mathrm{as}$. A minor part of the shift is due to the omission of the normalization term in \RGaia.

We show that fitting the parallaxes gives the same results as a comprehensive fit, thereby reducing the computational cost significantly compared to the latter. 

Our results show that the MW Cepheids may be fitted separately, with $b_W$ and $Z_W$ fixed to values provided from the remaining distance ladder, and the resulting value for $\MHW$ provided as an external constraint.

Fitting the MW Cepheids linearly gives results different from the comprehensive treatment and is therefore not recommended.

A color-dependent residual parallax correction can be motivated but exhibits only a minor $- 0.1 \, \mathrm{km/s/Mpc}$ additional shift in the Hubble constant.

\section{Dependence of $H_0$ on period}
\begin{figure}[t]
    \centering
    
    \includegraphics[width=\linewidth]{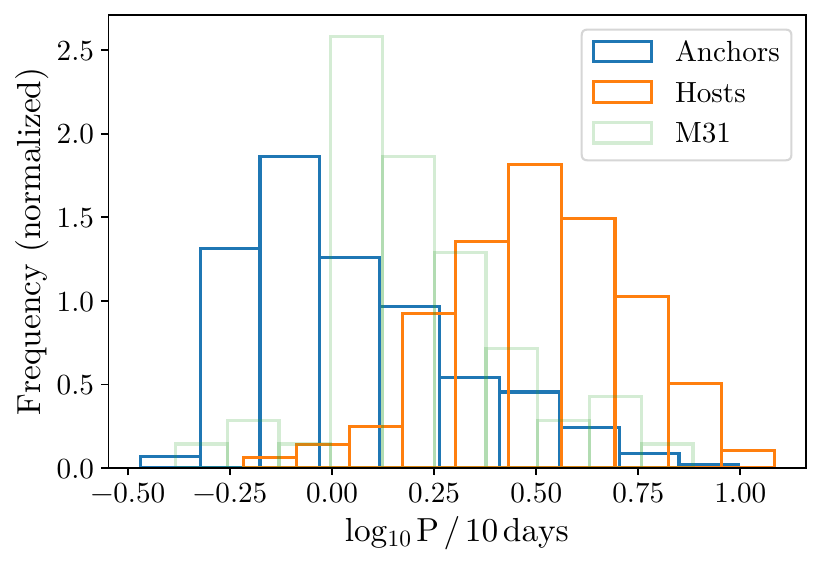}
    
    \caption{Normalized distribution of periods in anchor galaxies and host galaxies. The host galaxy periods are centered around higher periods than the anchor galaxies.
    \label{fig:PerDistr}}
\end{figure}

A potential source of systematic error in the Cepheid-based distance ladder is the difference in the distribution of Cepheid periods between anchor and host galaxy Cepheids. Fig.~\ref{fig:PerDistr} highlights these discrepancies, showing that Cepheids in anchor galaxies tend to have shorter periods compared to those in host galaxies. Such differences could bias the inferred Hubble constant if Cepheids with distinct periods exhibit different period-luminosity relations or belong to physically distinct populations.

To explore this effect, we analyze the inferred $H_0$ as a function of period. Specifically, we assess whether limiting the range of Cepheid periods in either the anchor or host galaxy populations, or both, affects the inferred value of $H_0$.
Additionally, we analyze a more flexible PLR, incorporating possible breaks at specific period thresholds, which can potentially mitigate any resulting bias. A simplified variant of this was analyzed in \Rbase with a fixed break at $\p = 0 \, \mathrm{dex}$ ($10 \, \mathrm{days}$). Here, we generalize and allow for several breaks and fit for their position as well as the slope in each period range.
Another method that we explore is to resample the Cepheids in a way which ensures that the period distribution agree between the anchor and host galaxies.
In the present section we utilize the same data sets as described in Section~\ref{sec:GlobalFit}, that is, the baseline \Rbase calibration plus recent updates. 

\subsection{Period-dependent trends in $H_0$}
\label{sec:PerTrends}
\begin{figure*}[t]
    \centering
    
    \includegraphics[width=0.485\linewidth]{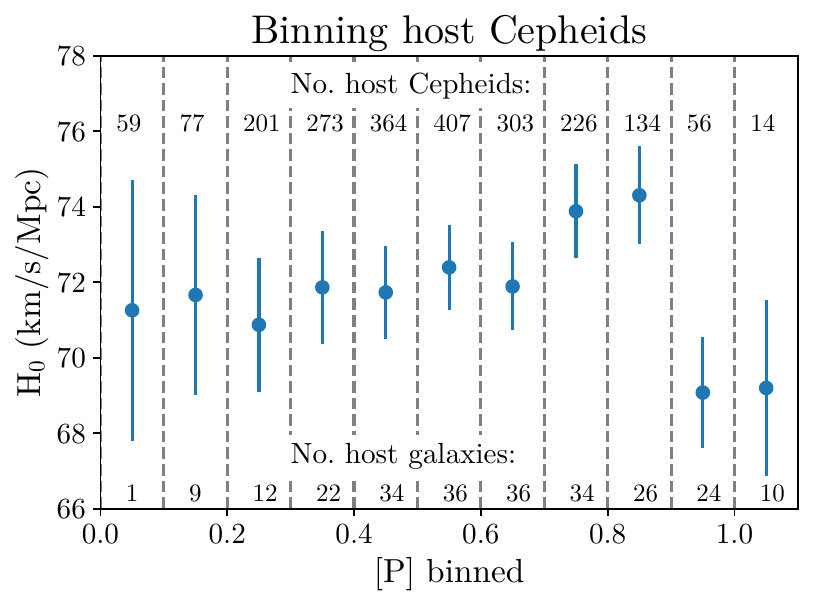}
    \includegraphics[width=0.495\linewidth]{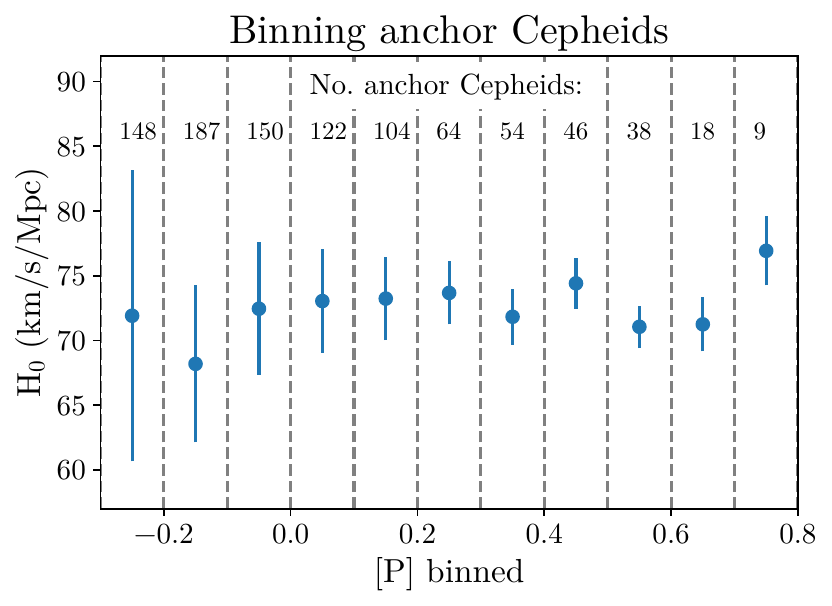}
    
    \caption{Dependence of $H_0$ on the Cepheid period. The number of Cepheids belonging to each bin is indicated on the top and the number of SNIa host galaxies with such Cepheids are indicated on the bottom. \emph{Left panel:} selecting host galaxy Cepheids with periods according to the bins, here of size $0.1$. All anchor galaxy Cepheids are retained. Host Cepheids in the range $\gtrsim 0.9$ give significantly lower values for $H_0$ compared with Cepheids of lower periods. For host Cepheids with $[P] \lesssim 0.9$, there is a clear correlation between decreasing period and decreasing $H_0$. \emph{Right panel:} selecting anchor Cepheids with periods according to the bins. All host galaxy Cepheids are retained. Here, for the anchor Cepheids, there is no obvious correlation between $H_0$ and period. 
    \label{fig:H0vsP}}
\end{figure*}

We begin by examining how the inferred $H_0$ varies when host galaxy Cepheids are included selectively based on their period. In this analysis, anchor galaxy Cepheids are retained in their entirety, while host Cepheids are grouped into bins of width $\Delta [P] = 0.1 \, \mathrm{dex}$. Fig.~\ref{fig:H0vsP} (left panel) shows that host Cepheids with longer periods ($[P] \gtrsim 0.9\, \mathrm{dex}$) yield significantly lower $H_0$ values than their shorter-period counterparts. Moreover, within the range $[P] \lesssim 0.9\, \mathrm{dex}$, a clear trend emerges: shorter-period Cepheids, that is, whose periods align more closely with those of the anchors, produce lower $H_0$ values.

This is similar to \cite{Freedman:2024eph} reporting a $3 \, \sigma$ significance for a positive correlation between the SNIa absolute magnitude, $M_B$, and the distance to the host galaxies. Since greater distance requires increased brightness to be observable, that is, increased period, this result points in the same direction, namely that host galaxy Cepheids which are, in some sense (either in distance or in period), closer to the anchor galaxy Cepheids yield a lower $H_0$ value.

In Appendix~\ref{sec:PerBinning}, we explore how $H_0$ depends on period when both anchor and host Cepheids are restricted to periods within a certain range.

Next, we repeat the analysis by limiting the period of the anchor Cepheids, while retaining all host Cepheids. Fig.~\ref{fig:H0vsP} (right panel) reveals no clear dependence of $H_0$ on the period of the anchor Cepheids when considering the full anchor sample.
However, we note that if the MW Cepheids are removed, there is a shift in $H_0$ at $[P] \sim 0.5 \, \mathrm{dex}$, akin to the host Cepheid behavior near $[P] \sim 0.9 \, \mathrm{dex}$. Anchor Cepheids with longer periods ($[P] \gtrsim 0.5 \, \mathrm{dex}$) yield noticeably lower $H_0$ values compared to those with shorter periods.

These results suggest that the difference in periods between anchor and host Cepheids might contribute to the inferred value of $H_0$.
In the following subsections, we explore a number of methods that aim at quantifying this effect.

\subsection{Matched-period resampling}
\label{sec:MatchedPerResamp}
One way of mitigating a potential bias in $H_0$ due to the difference in periods between anchors and hosts is to resample the Cepheids according to a specified, common distribution, referred to as the target distribution.

\begin{figure}[t]
    \centering
    
    \includegraphics[width=\linewidth]{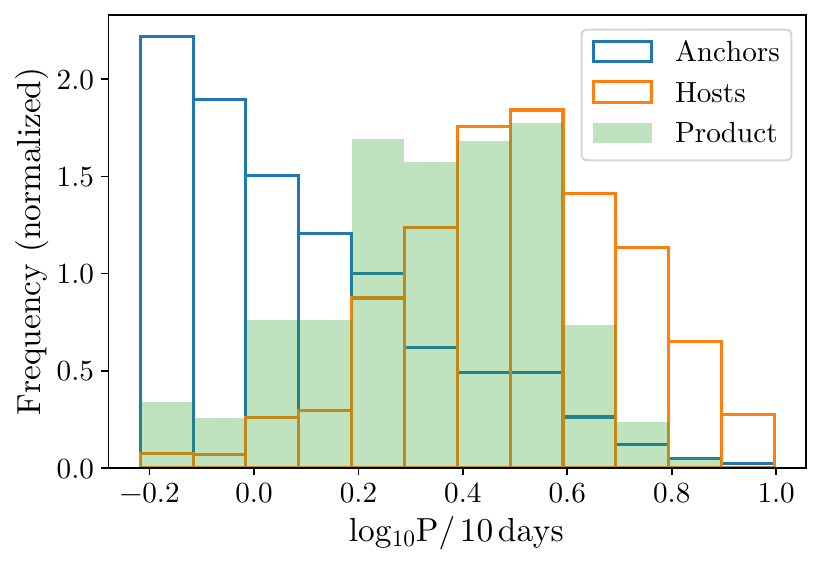}
    
    \caption{Normalized distribution of periods in anchor galaxies, host galaxies galaxies, and the product distribution of the two. Outside of the displayed window, the product distribution is zero.
    \label{fig:PerDistrProd}}
\end{figure}

Resampling was performed without replacement, using a target period distribution common to both anchor and host galaxies. This distribution was constructed as the product of their empirical period histograms (Fig.~\ref{fig:PerDistrProd}). In principle, the target distribution can be specified freely but we found empirically that the product distribution minimizes the increase in uncertainty due to resampling. The alternatives we have tested severely exacerbate the uncertainty, questioning their usefulness.

\begin{figure}[t]
    \centering
    
    \includegraphics[width=\linewidth]{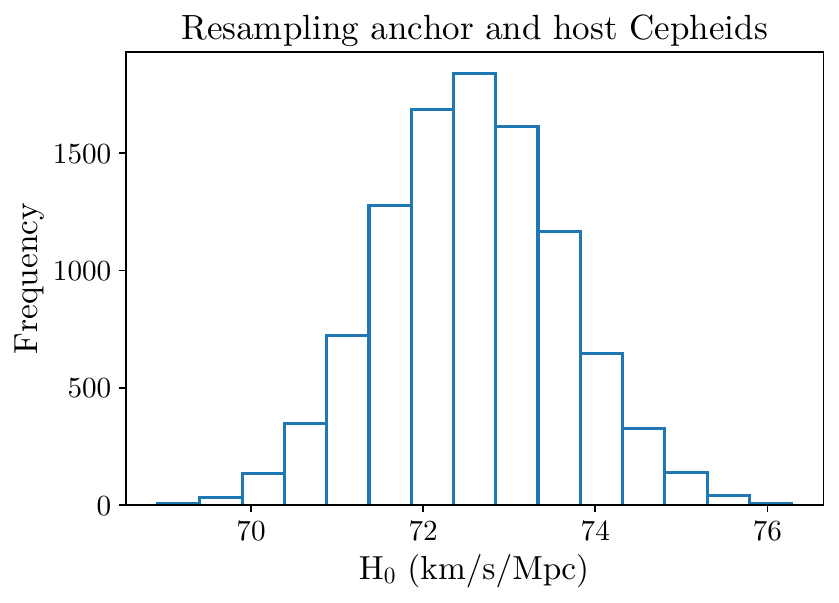}
    
    \caption{Sampled values of $H_0$ when the anchor and host galaxy Cepheids are resampled to follow a common period distribution. Due to computational costs, we make a linear fit of the MW Cepheids, so the reference value that should be compared with here, that is, keeping all Cepheids without resampling, is $H_0 = (73.05 \pm 0.90) \, \mathrm{km/s/Mpc}$.  The resampled values exhibit a mean value of $H_0 = 72.57 \, \mathrm{km/s/Mpc}$ with a mean uncertainty of $1.41 \, \mathrm{km/s/Mpc}$ and a mean dispersion due to the sampling of $1.05 \, \mathrm{km/s/Mpc}$. Thus, resampling the anchor and host Cepheids to enforce a consistent distribution of periods yields a $0.49 \, \mathrm{km/s/Mpc}$ decrease in the Hubble constant, according to this method.
    \label{fig:H0resamp}}
\end{figure}

Each iteration of the resampling randomly selects Cepheids, independently, for anchors and hosts based on the target distribution.
For each iteration, the Cepheid-based distance ladder was recalibrated to infer the Hubble constant. The process was repeated $10^4$ times, producing a robust distribution of $H_0$-values as shown in Fig.~\ref{fig:H0resamp}.
We fit the MW Cepheids simultaneously with the rest of the distance ladder but, due to computational limitations, the MW Cepheids are included linearly. As discussed previously, this leads to a slight overestimation of the Hubble constant but we still expect the relative change in $H_0$ to be reliably estimated in this way. For reference, when fitting the MW Cepheids linearly, without resampling (i.e. keeping all the Cepheids), we obtain $H_0 = (73.05 \pm 0.90) \, \mathrm{km/s/Mpc}$. 

The matched-period resampling results in a mean shift of
\begin{equation}
    \label{eq:DeltaH0PerSamp}
    \Delta H_0 = -0.49 \, \mathrm{km/s/Mpc}
\end{equation}
relative to the reference case where the full unresampled data was used. The distribution of $H_0$-values from resampled datasets is shown in Fig.~\ref{fig:H0resamp}.
While resampling mitigates the period difference, it comes at the cost of increased uncertainty. The mean uncertainty in $H_0$ among the resampled datasets is $1.41 \, \mathrm{km/s/Mpc}$ compared with $0.90 \, \mathrm{km/s/Mpc}$ if all Cepheids are used. This reflects the reduced sample size, with an approximate total of $700$ Cepheids after resampling (whereof $\simeq 220$ anchor Cepheids and $\simeq 480$ host Cepheids).
As shown in Fig.~\ref{fig:H0resamp}, there is an additional variability in $H_0$ due to the resampling. It reflects the systematic effect of choosing different Cepheids and represents an additional source of uncertainty. Here, the uncertainty due to the resampling is $1.05 \, \mathrm{km/s/Mpc}$.
The total uncertainty is obtained by adding the two contributions in quadrature to a total of $1.76 \, \mathrm{km/s/Mpc}$.

Adding the shift eq.~\eqref{eq:DeltaH0PerSamp} to eq.~\eqref{eq:H0_FullGlobal}, we obtain a final value of
\begin{equation}
    H_0 = (72.18 \pm 1.76) \, \mathrm{km/s/Mpc}.
\end{equation}
This is a conservative estimate that includes our renewed MW analysis as well as the resampling technique mitigating the difference in periods between anchors and hosts. 

The construction of the product distribution depends to some degree on the number of bins that we choose. The goal is to balance detail with stability, avoiding both over-smoothing (too few bins) and excessive noise (too many bins). Here, we have implemented Sturge's rule which is the default binning method in many data analysis softwares. With $3164$ data points, this results in $\simeq 12$ bins.
Varying the number of bins from 3 to 12 shows a consistent decrease in $H_0$ with $\Delta H_0$ ranging from $-0.63 \, \mathrm{km/s/Mpc}$ to $- 0.20 \, \mathrm{km/s/Mpc}$ and a mean value of $\Delta H_0 = - 0.38 \, \mathrm{km/s/Mpc}$.

\subsection{Broken PLR}
\label{sec:BrokenPLR}
The shift in $H_0$ at $\p \simeq 0.9 \, \mathrm{dex}$, and possibly at $\p \simeq 0.5 \, \mathrm{dex}$, discussed in Section~\ref{sec:PerTrends}, coupled with the trend of increasing $H_0$ with increasing host Cepheid period, calls into question the assumption of a single-linear PLR extending across almost two orders of magnitude in period.
To address this, we consider a generalized PLR model that allows for breaks at empirically determined period thresholds, enabling the PLR slope to vary across different ranges in period.

In principle, we can allow for an arbitrary number of breaks but we find empirically that the greatest increase in the quality of fit is obtained with two breaks, one at $\p_{b,1} = 0.465^{+0.015}_{-0.354} \, \mathrm{dex}$ and one at $\p_{b,2} = 0.909^{+0.020}_{-0.009} \, \mathrm{dex}$, as anticipated from Fig.~\ref{fig:H0vsP}. 
Within this model, the \Rbase variant with a break at $\p = 0 \, \mathrm{dex}$ is ruled out.

\begin{figure}[t]
    \centering
    
    \includegraphics[width=\linewidth]{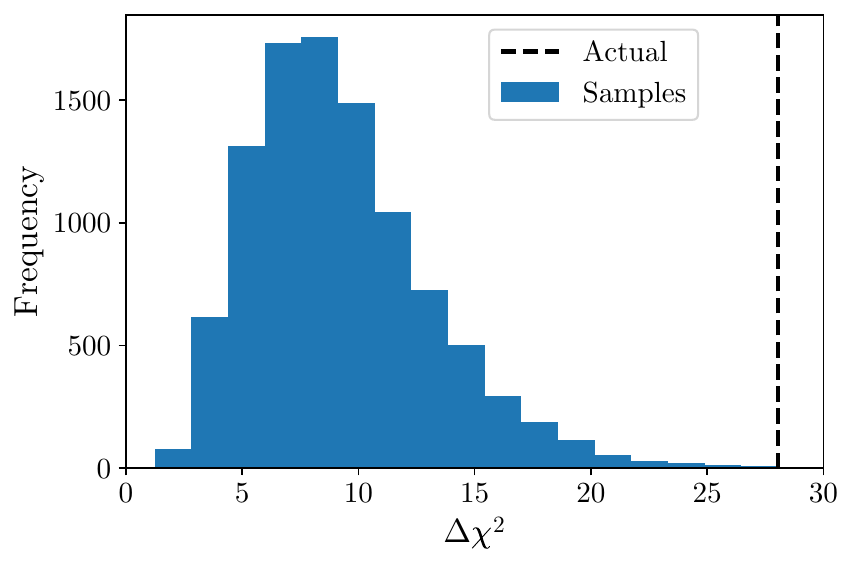}
    
    \caption{Distribution of $\Delta \chi^2$ values from 10,000 simulated datasets under the null hypothesis of a single-linear period-luminosity relation (PLR). The test statistic $\Delta \chi^2$ represents the improvement in fit achieved by the broken PLR compared to the single-linear PLR. The vertical dashed line at $\Delta \chi^2 = 28$ indicates the observed value from the actual data. Only $0.04 \, \%$ of simulations exceed this value ($p = 0.0004$), providing strong evidence to reject the null hypothesis in favor of the broken PLR.
    \label{fig:BrokenPLRSim}}
\end{figure}

The improvement in fit achieved by the broken PLR is quantified by a reduction in the $\chi^2$ value of 28 compared to the single-linear PLR. A null hypothesis test was conducted, with the single-linear PLR as the null hypothesis and the broken PLR as the alternative. Simulating 10,000 instances of the data under the null hypothesis revealed that a $\Delta \chi^2 \geq 28$ occurs in only $\simeq 0.04 \, \%$ of cases, corresponding to a p-value of $p = 0.0004$, see Fig.~\ref{fig:BrokenPLRSim}.\footnote{Due to the rare occurrence of these event where $\Delta \chi^2 \geq 28$, the p-value exhibits a degree of uncertainty due to the randomness in these simulations. Assuming the counts follow a Poisson distribution, we get $p = 0.0004 \pm 0.0002$. For simplicity, in the abstract and introduction, we have therefore instead stated an upper limit on the p-value as 0.001.}

To evaluate the relative parsimony and predictive utility of the two models, we employed the Akaike Information Criterion (AIC) and the Bayesian Information Criterion (BIC). The $\Delta$AIC value of $-20$ strongly favors the broken PLR, indicating that the improved fit outweighs the penalty for additional model complexity. By contrast, the $\Delta$BIC value of $+5$ slightly favors the single-linear PLR.

The discrepancy between the AIC and BIC arises from their differing penalties for model complexity. For large datasets, as in this analysis, BIC tends to favor simpler models more strongly than AIC. The $\Delta$BIC value thus reflects caution against overfitting, while the $\Delta$AIC value underscores the broken PLR’s improved predictive performance.

Here, we have used the fully updated SH0ES data, as described in Section~\ref{sec:GlobalFit}. Repeating the same exercise with only data from the fourth iteration of SH0ES (\Rbase) we get very similar results, with preferred breaking points around $0.47 \, \mathrm{dex}$ and $0.91 \, \mathrm{dex}$ with a $99.9 \, \%$ confidence rejection of the single-linear PLR in favour of the broken PLR (p-value $0.001$).

We have focused on the double-break PLR, which shows the most significant improvement in goodness of fit. Notably, introducing a single break in the PLR also improves the $\chi^2$ value, namely by 12 units. Conducting a hypothesis test for the single-break model against the single-linear PLR yields a p-value of 0.024. This result corresponds to a $97.6 \, \%$ confidence rejection of the single-linear PLR in favor of the single-break model.

Introducing a third break point improves the $\chi^2$-value by only 3 units compared to the two-break model. This represents a negligible improvement, and we therefore do not analyze this model further.

The breaking points split the Cepheids into three distinct populations, each with its own PLR slope: $b_W^\mathrm{short} = -3.257^{+0.034}_{-0.024} \, \mathrm{mag/dex}$ for the short-period Cepheids, $b_W^\mathrm{mid} = -3.347 \pm 0.020 \, \mathrm{mag/dex}$ for the mid-period Cepheids, and $b_W^\mathrm{long} = -3.236^{+0.033}_{-0.035} \, \mathrm{mag/dex}$ for the long-period Cepheids. 
Note the similarity between the slopes of the short-period and long-period Cepheids whereas the mid-period Cepheids have a significantly different slope, close to the theoretical expectation of $-10/3 \, \mathrm{mag/dex}$.\footnote{A PLR slope of $-2.5 \times 2 \times 2/3 = -10/3$ can be anticipated in the following way. The factor $-2.5$ is due to the relation between the magnitude and $\log L$ with $L$ being the luminosity. The factor $2$ is due to the exponent in the relation $L \propto R^2$, that is, Stefan--Boltzmann's law. The factor $2/3$ comes from $P \propto \rho^{-1/2}$, which is a well-known relation for mechanical systems, here the Cepheids, together with $\rho \propto R^{-3}$. See e.g. \cite{Madore:1991yf}.}

\begin{figure*}[t]
    \centering
    
    \includegraphics[width=\linewidth]{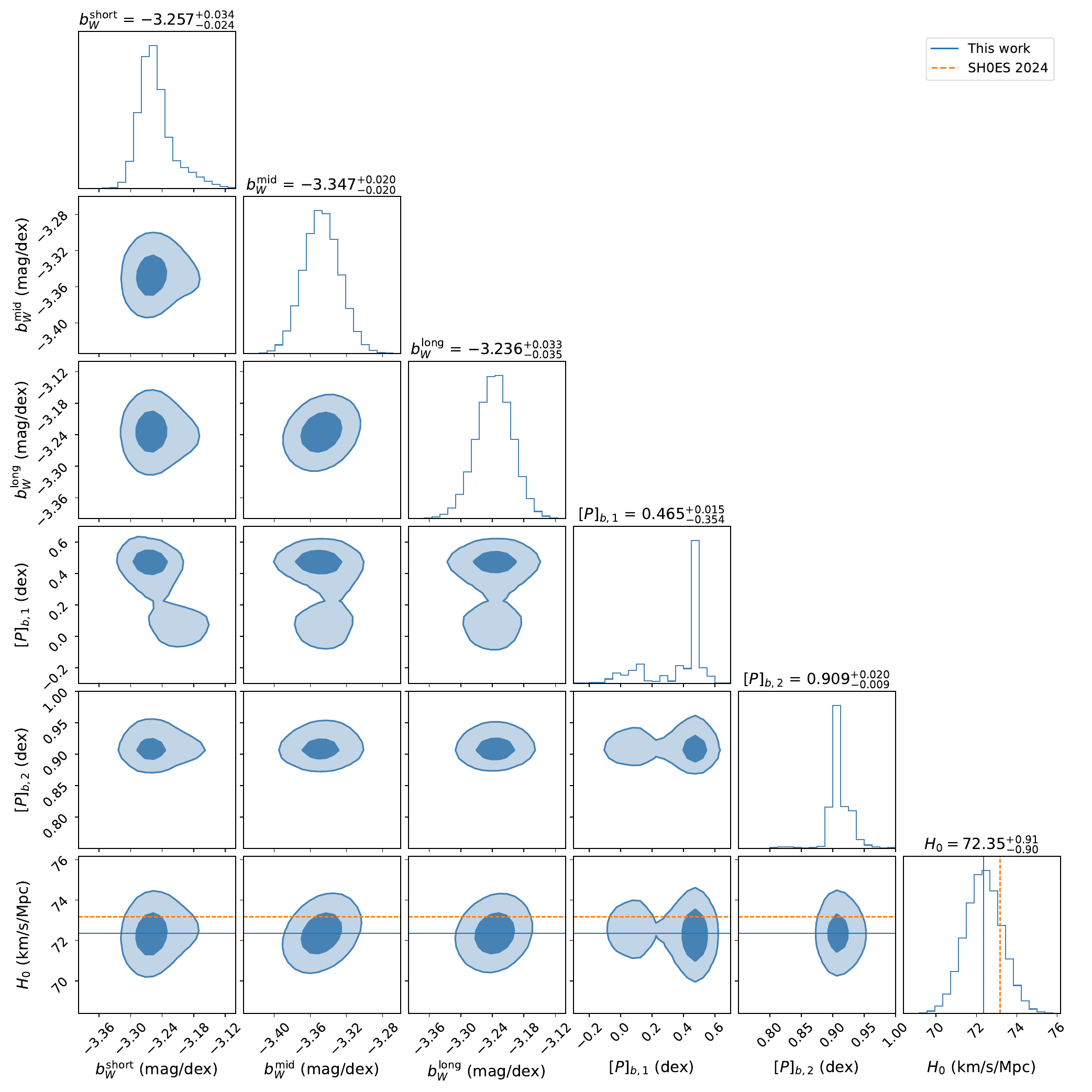}
    
    \caption{Marginalized posterior distributions for a selection of model parameters when the distance ladder is calibrated with a double-break PLR. Shaded regions indicate the $39 \, \%$ and $86 \, \%$ credible intervals, corresponding to the $1 \, \sigma$ and $2 \, \sigma$ confidence levels, respectively. There is a preference for a first break at $\p_{b,1} \simeq 0.47 \, \mathrm{dex}$ and clear preference for a second break at $\p_{b,2} \simeq 0.91 \, \mathrm{dex}$. Note the slight bimodality of $\p_{b,1}$ with a secondary mode at $\simeq 0.1 \, \mathrm{dex}$.
    The PLR slopes in the low-period and high-period regions are similar $b_W^\mathrm{short},b_W^\mathrm{long} \simeq -3.2 \, \mathrm{dex}$ whereas the mid-period slope, with $b_W^\mathrm{mid} = (-3.35 \pm 0.02) \, \mathrm{dex}$, is close to the theoretical expectation. With a broken PLR, the $H_0$-value decreases by $0.32 \, \mathrm{km/s/Mpc}$ compared with a single-linear PLR model for a total shift of $- 0.82 \, \mathrm{km/s/Mpc}$ compared with the SH0ES team \citep{Breuval:2024lsv}.
    \label{fig:PLRbreak}}
\end{figure*}

In Fig.~\ref{fig:PLRbreak}, we show the 2d marginalized posterior distributions for a model allowing two breaks in the PLR. Note that the location of the breaks, as well as the PLR slopes, are model parameters that we fit and that we include the MW Cepheids in a global fit, as in Section~\ref{sec:GlobalFit}.

The inferred value for the Hubble constant is
\begin{equation}
    \label{eq:H0PLRbreak}
    H_0 = (72.35 \pm 0.91) \, \mathrm{km/s/Mpc}.
\end{equation}
Compared with eq.~\eqref{eq:H0_FullGlobal}, that is, the value obtained when making the same fit without any breaks in the PLR we see that allowing for breaks in the PLR decreases the $H_0$-value by $\simeq 0.32 \, \mathrm{km/s/Mpc}$ for a total shift of $- 0.82 \, \mathrm{km/s/Mpc}$, when including the MW recalibration, compared with the SH0ES team \citep{Breuval:2024lsv}.
From Fig.~\ref{fig:PLRbreak}, it is evident that the posterior distribution is non-Gaussian with a slight bimodality in $\p_{b,1}$ with a secondary mode at $\simeq 0.1 \, \mathrm{dex}$. However, the figure indicates that this feature does not affect the Hubble constant.
We also note that, while the median and mean of $H_0$ agree to the value in eq.~\eqref{eq:H0PLRbreak}, the mode (maximum likelihood) occurs at a lower value of $H_0 = 71.62 \, \mathrm{km/s/Mpc}$. We attribute this to the non-Gaussianity of the posterior distribution.

\section{Discussion and Conclusions}
\label{sec:Discussion}
We have revisited the calibration of the Cepheid-based distance ladder, exploring two potential sources of systematic uncertainties: (1) The difference in the distribution of Cepheid periods between anchor and SNIa host galaxies. (2) The assumed prior on the residual parallax offset of the (MW) Cepheids. The results provide insights into the robustness of current Hubble constant estimates and highlight areas for improvement in methodology.

The SH0ES team \cite{Riess_2021} (\RGaia) choose a prior on the residual parallax offset, $zp$, which is centered around zero with a standard deviation of $10 \, \mu \mathrm{as}$. Considering the uncertainty in the parallax correction, as highlighted in \cite{Lindegren_2021a,Lindegren_2021b}, and the slight inconsistencies between some of the external calibrations \citep{Bhardwaj:2021,Zinn:2021,Stassun:2021,Fabricius:2021,Huang:2021}, we follow a more conservative approach and impose an uninformative prior on $zp$.
In this way, we obtain an independent external constraint of $zp = -16 \pm 6 \, \mu \mathrm{as}$ in the Cepheid magnitude range, when calibrating the full distance ladder.

Our reanalysis of the MW Cepheids results in a fiducial magnitude for the MW Cepheids which is $1.0 \, \sigma$ lower than that of \RGaia as well as a more significant evidence for a negative residual parallax offset in the Gaia EDR3 parallax measurements; $> 3 \, \sigma$ compared with $2 \, \sigma$ in \RGaia.
In the full distance ladder this leads to a $-0.6 \, \mathrm{km/s/Mpc}$ shift in the Hubble constant compared with the corresponding SH0ES results ($-0.5 \, \mathrm{km/s/Mpc}$ with the full updated data sets), or $-1.4 \, \mathrm{km/s/Mpc}$ if the MW is used as the sole anchor.

We show that treating the MW Cepheids separately, but with fixed PLR slopes ($b_W$ and $Z_W$), using the inferred fiducial Cepheid magnitude as an external constraint in the remaining distance ladder, yields results consistent with a full global fit, providing a computationally efficient alternative for future analyses.

We also analyze the effects of the differing period distribution between the anchor and SNIa host galaxies, which introduces a potential bias in 
$H_0$. Anchor Cepheids exhibit shorter periods on average, while SNIa host Cepheids skew towards longer periods. This discrepancy affects the inferred 
$H_0$ value, as demonstrated by the trend of decreasing 
$H_0$ when excluding longer-period Cepheids from the host population.

To address the period difference, we introduced a resampling technique that aligns the anchor and host Cepheid distributions to a common period distribution. This method reduces the bias, leading to a $-0.49 \, \mathrm{km/s/Mpc}$ shift in $H_0$.
At the same time, the resampling approach increases uncertainty due to reduced sample size, with a total resulting uncertainty of $1.76 \, \mathrm{km/s/Mpc}$, underscoring the trade-offs inherent in mitigating systematic biases.
Together with the renewed MW analysis, this leads to a final value $H_0 = (72.18 \pm 1.76) \, \mathrm{km/s/Mpc}$ which is a total $-0.99 \, \mathrm{km/s/Mpc}$ shift in $H_0$ and a doubling of the uncertainty.
This is a conservative estimate of the Hubble constant that mitigates the difference in Cepheid periods between the anchor and SNIa host galaxies.
Together, this results in a reduction of the Hubble tension from $5.4 \, \sigma$ to $2.4 \, \sigma$. 

Another method that can potentially address the systematic effects of period difference between anchors and hosts, is a more flexible alternative to the single-linear period-luminosity relation (PLR).
Specifically, we consider a double-break PLR, one of the simplest extensions, with the aim of capturing variations in Cepheid populations spanning nearly two orders of magnitude in period.
Within this model, we see a preference for breaks in the PLR at $0.47 \, \mathrm{dex}$ and $0.91 \, \mathrm{dex}$ and a significant improvement in the quality of fit.
The high significance of the p-value for the double-break PLR against the single-linear PLR ($p = 0.0004$) and the $\Delta$AIC value both support the broken PLR as the preferred model. The BIC’s slight preference for the single-linear PLR reflects its stringent penalty for additional parameters.

The introduction of the broken PLR was not guided by any specific theoretical considerations regarding the physical properties of Cepheids but rather motivated by the improvement it provides in describing the observed data.
One possible explanation for the improved fit provided by the broken PLR is the younger age of long-period Cepheids, which are more likely to be found in open clusters \citep{Anderson:2016txx}. This association could introduce biases in the photometry of these Cepheids, as standard crowding corrections might fail to fully account for the additional light from unresolved cluster members. 
A generalized PLR might be able to better mitigate this effect. 
This is just one example of the systematic effects that might arise when calibrating the distance ladder using short-period Cepheids in the first rung and long-period Cepheids in the second rung. A more extensive discussion can be found in \cite{Kushnir:2024spm}.

The broken PLR model yields a $-0.32 \, \mathrm{km/s/Mpc}$ shift in the Hubble constant compared with the single-linear PLR.
Together with the renewed MW analysis, this yields a final value $H_0 = (72.35 \pm 0.91) \, \mathrm{km/s/Mpc}$, corresponding to a reduction in the Hubble tension from $5.4 \, \sigma$ to $4.4 \, \sigma$.

In summary, we have explored two potential sources of systematic effects in the Cepheid-based distance ladder that bias the Hubble constant high---systematic differences in the Cepheid periods between anchor and SNIa host galaxies and the influence of the residual parallax offset of the MW Cepheids, resulting in a total $ \simeq -1 \, \mathrm{km/s/Mpc}$ shift in the Hubble constant.
Our findings underscore the importance of careful consideration of Cepheid population characteristics in $H_0$ calibrations.

\begin{acknowledgements}
Thanks to an anonymous referee for reading the manuscript.
Thanks to Yukei Murakami for helpful discussions concerning the MCMC code of \cite{Riess:2021jrx}, publicly available on GitHub\footnote{\href{https://github.com/PantheonPlusSH0ES/DataRelease/tree/main/SH0ES_Data}{\url{https://github.com/PantheonPlusSH0ES/DataRelease/tree/main/SH0ES_Data}} (last checked 2024-03-27)} and for providing the updated SN covariance matrix of \cite{Murakami:2023xuy}.
We thank Adam Riess for helpful explanations and discussions and for providing data for the Cepheid (Z-Sct) which is missing in the tables of \cite{Riess_2021}. We are also grateful for his careful reading of the first version of the manuscript, providing several useful comments and insights.

MH and EM acknowledges support from the Swedish Research Council under Dnr VR 2020-03384.

This research utilized the Sunrise HPC facility supported by the Technical Division at the Department of Physics, Stockholm University.

The authors used OpenAI's ChatGPT to assist in drafting portions of the following sections: Abstract, Introduction, and Conclusions.
\end{acknowledgements}

\section*{Data availability}
The data used in this study, pertaining to the fourth iteration of the SH0ES team calibration, \Rbase, are publicly available on GitHub at \href{https://github.com/marcushogas/Cepheid-Distance-Ladder-Data}{\url{https://github.com/marcushogas/Cepheid-Distance-Ladder-Data}}. These data are freely accessible for use and redistribution under the terms of the repository.

Data related to the MW Cepheids from \RGaia are publicly available at \href{https://dx.doi.org/10.3847/2041-8213/abdbaf}{\url{https://dx.doi.org/10.3847/2041-8213/abdbaf}}.

The MW Cepheid data from the Cepheid cluster analysis of \cite{Riess:2022mme} can be accessed at \href{https://dx.doi.org/10.3847/1538-4357/ac8f24}{\url{https://dx.doi.org/10.3847/1538-4357/ac8f24}}.

Updated data for the SMC Cepheids from \cite{Breuval:2024lsv} are publicly available at \href{https://dx.doi.org/10.3847/1538-4357/ad630e}{\url{https://dx.doi.org/10.3847/1538-4357/ad630e}}.

The updated SNIa covariance matrix from \cite{Murakami:2023xuy} is owned by the authors and is not publicly available. The data were made available upon request to the authors.

\bibliographystyle{mnras}
\bibliography{bibliography}{}

\appendix

\section{Simplified MW likelihood}
\label{sec:like_parallax}
A simplification of the likelihood in eq.~\eqref{eq:like_full} can be achieved in the following way. The likelihood can be written
\begin{widetext}
    \begin{equation}
    \label{eq:like_interm}
    \mathcal{L}(\theta) = \prod_{i=1}^{66} \frac{\exp \left[ - \frac{1}{2} \left(\frac{m^W_{H,i} - m^{W,\mathrm{obs}}_{H,i}}{\sigma_{m,i}} \right)^2 
 - \frac{1}{2} \left(  \frac{\pi_{i}(\theta) - \pi_{\mathrm{EDR3},i}}{\sigma_{\pi,i}} \right)^2 \right]}{\sqrt{2\pi \sigma^2_{m,i} \times 2\pi \sigma^2_{\pi,i}}} .
\end{equation}
\end{widetext}
Here, we let $(m_{H,i}^W, \MHW, b_W, Z_W, zp)$ be the model parameters. We reparameterize from $m_{H,i}^W$ to $\delta m_{H,i}^W$, defined via,
\begin{equation}
    m_{H,i}^W = m_{H,i}^{W,\mathrm{obs}} + \delta m_{H,i}^W. 
\end{equation}
Assuming that the model provides a good fit to data we have $\delta m_{H,i}^W \ll m_{H,i}^W$. Taylor expanding $\pi_i(\theta)$ around $\delta m_{H,i}^W / m_{H,i}^W = 0$ results in
\begin{widetext}
    \begin{equation}
    \label{eq:like_interm2}
    \mathcal{L}(\theta) = \prod_{i=1}^{66} \frac{\exp \left[ - \frac{1}{2} \left(\frac{\delta m^W_{H,i}}{\sigma_{m,i}} \right)^2 
 - \frac{1}{2} \left(  \frac{\pi_{i}^\mathrm{phot}(\theta) + \left. \left( \frac{\partial \pi_i}{\partial m_{H,i}^W} \right) \right|_{m_{H,i}^{W,\mathrm{obs}}} \delta m_{H,i}^W + \mathcal{O}((\delta m_{H,i}^W)^2) - \pi_{\mathrm{EDR3},i}}{\sigma_{\pi,i}} \right)^2 \right]}{\sqrt{2\pi \sigma^2_{m,i} \times 2\pi \sigma^2_{\pi,i}}} .
\end{equation}
\end{widetext}
Ignoring the higher-order terms $\mathcal{O}((\delta m_{H,i}^W)^2)$ and marginalizing over $\delta m_{H,i}^W$ gives
\begin{equation}
    \label{eq:like_parallax_2}
    \mathcal{L}(\theta) = \prod_{i=1}^{66} \frac{\exp \left[ - \frac{1}{2} \left(  \frac{\pi_{i}^\mathrm{phot}(\theta) - \pi_{\mathrm{EDR3},i}}{\sigma_{\mathrm{tot},i}} \right)^2 \right]}{\sqrt{ 2\pi \sigma^2_{\mathrm{tot},i}}},
\end{equation}
where 
\begin{subequations}
    \begin{align}
        \pi_i^\mathrm{phot}(\theta) &= 10^{(10 - \mu_i^\mathrm{phot}(\theta)) / 5} - zp,\\
        \mu_i^\mathrm{phot}(\theta) &= m_{H,i}^{W,\mathrm{obs}} - \MHW - b_W \p_i - Z_W \oh_i,
    \end{align}
\end{subequations}
and the total error $\sigma_{\mathrm{tot},i}$ includes the error in the parallax as well as in the magnitude, that is,
\begin{equation}
    \sigma_{\mathrm{tot},i}^2 = \sigma_{\pi, i}^2  + \left( 10^{-\mu_i^\mathrm{phot}(\theta)/5} \, 20 \ln 10 \, \sigma_{m,i} \right)^2 .
\end{equation}
Note that the prefactor of $\sigma_{m,i}$ is $\left. \left( \frac{\partial \pi_i}{\partial m_{H,i}^W} \right) \right|_{m_{H,i}^{W,\mathrm{obs}}}$.
We see that $\sigma_{\mathrm{tot},i}$ depends on the model parameters via $\mu_i^\mathrm{phot}(\theta)$. This highlights the importance of the normalization term in eq.~\eqref{eq:like_parallax_2} and eq.~\eqref{eq:like_parallax}.

\section{Global fit likelihood}
\label{sec:GlobalFitLikelihood}
Here, we give a comprehensive account of the likelihood used in the global fit of Section~\ref{sec:GlobalFit}, fitting the MW Cepheids together with the rest of the distance ladder rather than providing them as external constraints and using the most up-to-date data. 

We split the contributions to the log-likelihood into four terms, one for the Gaia EDR3 MW Cepheids, one for the MW cluster Cepheids, one for the HST MW Cepheids, and one for the rest of the distance ladder,
\begin{align}
    \ln \mathcal{L}_\mathrm{tot} = &\ln \mathcal{L}_\mathrm{MW \; Gaia} + \ln \mathcal{L}_\mathrm{MW \; cluster} \nonumber \\  
    &+ \ln \mathcal{L}_\mathrm{MW \; HST} + \ln \mathcal{L}_\mathrm{rest} .
\end{align}
We use 66 MW Cepheids with parallaxes from Gaia EDR3 and the likelihood
\begin{equation}
    \label{eq:like_Gaia}
    \mathcal{L}_\mathrm{MW \; Gaia} = \prod_{i=1}^{66} \frac{\exp \left[ - \frac{1}{2} \left(  \frac{\pi_{i}^\mathrm{phot} - \pi_{\mathrm{EDR3},i}}{\sigma_{\mathrm{tot},i}} \right)^2 \right]}{\sqrt{ 2\pi \sigma^2_{\mathrm{tot},i}}},
\end{equation}
where 
\begin{subequations}
    \begin{align}
        \pi_i^\mathrm{phot} &= 10^{(10 - \mu_i^\mathrm{phot}) / 5} - zp_\mathrm{Gaia},\\
        \mu_i^\mathrm{phot} &= m_{H,i}^{W,\mathrm{obs}} - \MHW - b_W \p_i - Z_W \oh_i,
    \end{align}
\end{subequations}
and the total uncertainty $\sigma_{\mathrm{tot},i}$ includes the error in the parallaxes, magnitudes, and metallicities, that is,
\begin{align}
    \label{eq:sigma_tot}
    \sigma_{\mathrm{tot},i}^2 = \sigma_{\pi, i}^2  &+ \left( 10^{-\mu_i^\mathrm{phot}/5} \, 20 \ln 10 \, \sigma_{m,i} \right)^2 \nonumber \\
    & + \left( 10^{-\mu_i^\mathrm{phot}/5} \, 20 \ln 10 \, Z_W \, \sigma_{\oh,i} \right)^2.
\end{align}
Recall that $\sigma_{m,i}$ includes contributions from the photometry and the intrinsic scatter, added in quadrature. 

For the MW Cepheids with parallaxes from open clusters, the likelihood is similar to eq.~\eqref{eq:like_Gaia} with,
\begin{equation}
    \label{eq:like_cluster}
    \mathcal{L}_\mathrm{MW \; cluster} = \prod_{i=1}^{17} \frac{\exp \left[ - \frac{1}{2} \left(  \frac{\pi_{i}^\mathrm{phot} - \pi_{\mathrm{cluster},i}}{\sigma_{\mathrm{tot},i}} \right)^2 \right]}{\sqrt{ 2\pi \sigma^2_{\mathrm{tot},i}}},
\end{equation}
where 
\begin{subequations}
    \begin{align}
        \pi_i^\mathrm{phot} &= 10^{(10 - \mu_i^\mathrm{phot}) / 5} - zp_\mathrm{cluster},\\
        \mu_i^\mathrm{phot} &= m_{H,i}^{W,\mathrm{obs}} - \MHW - b_W \p_i - Z_W \oh_i,
    \end{align}
\end{subequations}
and the total error $\sigma_{\mathrm{tot},i}$ has the same form as in eq.~\eqref{eq:sigma_tot}.

The likelihood for the MW Cepheids with HST parallaxes assumes a similar form but without the residual parallax offset,
\begin{equation}
    \label{eq:like_HST}
    \mathcal{L}_\mathrm{MW \; HST} = \prod_{i=1}^{7} \frac{\exp \left[ - \frac{1}{2} \left(  \frac{\pi_{i}^\mathrm{phot} - \pi_{\mathrm{HST},i}}{\sigma_{\mathrm{tot},i}} \right)^2 \right]}{\sqrt{ 2\pi \sigma^2_{\mathrm{tot},i}}},
\end{equation}
where 
\begin{subequations}
    \begin{align}
        \pi_i^\mathrm{phot} &= 10^{(10 - \mu_i^\mathrm{phot}) / 5},\\
        \mu_i^\mathrm{phot} &= m_{H,i}^{W,\mathrm{obs}} - \MHW - b_W \p_i - Z_W \oh_i,
    \end{align}
\end{subequations}
and the total error $\sigma_{\mathrm{tot},i}$ has the same form as in eq.~\eqref{eq:sigma_tot}.

The likelihood for the remaining distance ladder reads
\begin{equation}
    \mathcal{L}_\mathrm{rest} = e^{- \frac{1}{2} (y-Lq)^T C^{-1} (y-Lq)}.
\end{equation}
Here, the likelihood is unnormalized, since the covariance matrix is constant and does not depend on the model parameters. The data vector $(y)$, design matrix $(L)$, covariance matrix $(C)$, and the vector of model parameters $(q)$ take the following form:
\begin{widetext}
    \begin{equation}
    \label{eq:y_update}
    y = 
    \begin{array}{ll}
    \left(
    \begin{array}[c]{c}
    
    m^W_{H,\mathrm{M101}} \\

    : \\

    m^W_{H,\mathrm{U9391}} \\
    
    \hline
    
    m^W_{H,\textrm{N4258}} \\
    
    m^W_{H,\textrm{M31}} \\
    
    m^W_{H,\textrm{LMC,GRND}} \\

    m^W_{H,\textrm{LMC,HST}} \\
    
    m^W_{H,\textrm{SMC}} \\
    
    \hline
    
    m_{B,\mathrm{Cal \; SN}} \\

    \hline
    
    m_{B,\mathrm{HF \; SN}} - 5 \log c z \{ ... \} -25 \\
    
    \hline
    
    0 \\
    
    \mu_\mathrm{N4258}^\mathrm{anch} \\
    
    \mu_\mathrm{LMC}^\mathrm{anch}\\

    \Delta \mu_\mathrm{SMC/LMC}^\mathrm{anch}
    \end{array} \right)
    
    &
    
    \begin{array}[c]{@{}l@{\,}l}
    
    \left.
    \begin{array}{c} \vphantom{m^W_{H,\mathrm{hosts}}} \\
    \vphantom{:} \\
    \vphantom{m^W_{H,\mathrm{hosts}}}
    \end{array}
    \right\} & \text{2150 Cepheids in SNIa hosts} \\
    
    \left.
    \begin{array}{c} \vphantom{m^W_{H,\textrm{nh},j}} \\ 
    \vphantom{m^W_{H,\textrm{nh},j}} \\ 
    \vphantom{m^W_{H,\textrm{nh},j}} \\
    \vphantom{m^W_{H,\textrm{nh},j}} \\
    \vphantom{\textrm{\LARGE HELLO}}
    \end{array}
    \right\} & \text{(443 + 55 + 270 + 69 + 87) Cepheids in anchors or non-SNIa hosts\hspace{1.5in}} \\
    
    \left.
    \begin{array}{c}
    \vphantom{m_B^0}
    \end{array}
    \right\} & \text{77 Cal SNe magnitudes} \\

    \left.
    \begin{array}{c}
    \vphantom{.}
    \end{array}
    \right\} & \text{277 HF SNe} \\
    
    \left.
    \begin{array}{c}
    \vphantom{0} \\
    \vphantom{0} \\
    \vphantom{0} \\
    \vphantom{\textrm{\LARGE HELLO}}
    \end{array}
    \right\} & \text{4 External constraints}
    \end{array}
    \end{array}
    \end{equation}

    \begin{equation}
    \label{eq:C_update}
        C = \left(
        \begin{array}{ccccccccccccccc}
        
        \sy{\sigma_{\rm M101}^2}\!\!\!\! & \td & \sy{Z_{\textrm{cov}}} & \sy{Z_{\textrm{cov}}} & \sy{0} & \sy{0} & \sy{0} & \sy{0} & \sy{0} & \sy{0}  & \sy{0} & \sy{0} & \sy{0} & \sy{0} \\
        
        : & \rotatebox{45}{:} & : & : & : & : & : & : & : & : & : & : & : & : \\
        
        \sy{Z_{\textrm{cov}}} & \td & \sy{\sigma_{\rm U9391}^2}\!\!\!\! & \sy{Z_{\textrm{cov}}} & \sy{0} & \sy{0} & \sy{0} & \sy{0} &\sy{0} & \sy{0} & \sy{0} & \sy{0} & \sy{0} & \sy{0} \\
        
        \hline
        
        \sy{Z_{\textrm{cov}}} & \td & \sy{Z_{\textrm{cov}}} & \sy{\sigma_{{\rm N4258}}^2}\!\!\!\! & \sy{0} & \sy{0} & \sy{0} & \sy{0}  &\sy{0} & \sy{0} & \sy{0} & \sy{0} & \sy{0} & \sy{0} \\
        
        \sy{0} & \td & \sy{0} & \sy{0} & \sy{\sigma_{{\rm M31}}^2}\!\!\!\! & \sy{0} & \sy{0} & \sy{0}  &\sy{0} & \sy{0} & \sy{0} & \sy{0} & \sy{0} & \sy{0} \\

        \sy{0} & \td & \sy{0} & \sy{0} & \sy{0} &  \sy{\sigma_{{\rm LMC,GRND}}^2}\!\!\!\! & \sy{10^{-4}} & \sy{0} & \sy{0} & \sy{0} & \sy{0} & \sy{0} & \sy{0} & \sy{0} \\

        \sy{0} & \td & \sy{0} & \sy{0} & \sy{0} & \sy{10^{-4}} &  \sy{\sigma_{{\rm LMC,HST}}^2}\!\!\!\! & \sy{0} & \sy{0} & \sy{0} & \sy{0} & \sy{0} & \sy{0} & \sy{0} \\

        \sy{0} & \td & \sy{0} & \sy{0} & \sy{0} & \sy{0} & \sy{0} &  \sy{\sigma_{{\rm SMC}}^2}\!\!\!\! & \sy{0} & \sy{0} & \sy{0} & \sy{0} & \sy{0} & \sy{0} \\

        \hline
                
        \sy{0} & \td & \sy{0} & \sy{0} & \sy{0} & \sy{0} & \sy{0} & \sy{0} & \sy{\sigma^2_{\textrm{Cal SN}}}\!\!\!\! & \sy{{\rm SN}_{\textrm{cov}}} & \sy{0} & \sy{0} & \sy{0} & \sy{0} \\
        
        \hline

        \sy{0} & \td & \sy{0} & \sy{0} & \sy{0} & \sy{0} & \sy{0} & \sy{0} & \sy{{\rm SN}_{\textrm{cov}}} & \sy{\sigma^2_{\textsc{HF SN}}}\!\!\!\! & \sy{0} & \sy{0} & \sy{0} & \sy{0} \\
        
        \hline
                
        \sy{0} & \td & \sy{0} & \sy{0} & \sy{0} & \sy{0} & \sy{0} & \sy{0} & \sy{0} & \sy{0} & \sy{\sigma_{\rm grnd}^2}\!\!\!\! & \sy{0} & \sy{0} & \sy{0} \\

        \sy{0} & \td & \sy{0} & \sy{0} & \sy{0} & \sy{0} & \sy{0} & \sy{0} & \sy{0} & \sy{0} & \sy{0} & \sy{\sigma_{\mu,{\rm N4258}}^2}\!\!\!\! & \sy{0} & \sy{0} \\
        
        \sy{0} & \td & \sy{0} & \sy{0} & \sy{0} & \sy{0} &  \sy{0} & \sy{0} & \sy{0} & \sy{0} & \sy{0} & \sy{0} & \sy{\sigma_{\mu,{\rm LMC}}^2} & \sy{0} \\
        
        \sy{0} & \td & \sy{0} & \sy{0} & \sy{0} & \sy{0} &  \sy{0} & \sy{0} & \sy{0} & \sy{0} & \sy{0} & \sy{0} & \sy{0} & \sy{\sigma_{\Delta \mu,{\rm SMC/LMC}}^2}
        \end{array}
        \right)
    \end{equation}

    \begin{equation}
    \label{eq:Lq_update}
        L =
        \left(
        \begin{array}[c]{ccccccclclccc}
        
        1 & \td & 0 & 0 & 1 & 0 & 0 & [P]_\mathrm{M101} & 0 & \oh_\mathrm{M101} & 0 & 0 & 0 \\
        
        : & \rotatebox{45}{:} & : & : & : & : & :   & : & : & : & : & : & : \\  
        
        0 & \td & 1 & 0 & 1 & 0 & 0 & [P]_\mathrm{U9391} & 0 & \oh_\mathrm{U9391} & 0 & 0 & 0 \\
        
        \hline
        
        0 & \td & 0 & 1 & 1 & 0 & 0 & [P]_\mathrm{N4258} & 0 & \oh_{\textrm{N4258}} & 0 & 0 & 0 \\
        
        0 &\td & 0 & 0 & 1 & 0 & 1 & [P]_\mathrm{M31} & 0 & \oh_{\textrm{M31}}   & 0 & 0 & 0 \\
        
        0 & \td & 0 & 0 & 1 & 1 & 0 & [P]_\mathrm{LMC,GRND} & 0 & \oh_{\textrm{LMC,GRND}} & 1 & 0 & 0 \\

        0 & \td & 0 & 0 & 1 & 1 & 0 & [P]_\mathrm{LMC,HST} & 0 & \oh_{\textrm{LMC,HST}} & 0 & 0 & 0 \\

        0 & \td & 0 & 0 & 1 & 1 & 0 & [P]_\mathrm{SMC} & 0 & \oh_{\textrm{SMC}} & 0 & 0 & 1 \\
        
        \hline
        
        1 & \td & 0 & 0 & 0 & 0 & 0 & 0 & 1 & 0 & 0 & 0 & 0 \\
        
        : & \rotatebox{45}{:} & : & : & : & : & : & : & : & : & : & : & : \\  
        
        0 & \td & 1 & 0 & 0 & 0 & 0 & 0 & 1 & 0 & 0 & 0 & 0 \\
        
        \hline

        0 & \td & 0 & 0 & 0 & 0 & 0 & 0 & 1 & 0 & 0 & -1 & 0 \\

        : & \rotatebox{45}{:} & : & : & : & : & : & : & : & : & : & : & : \\
        
        0 & \td & 0 & 0 & 0 & 0 & 0 & 0 & 1 & 0 & 0 & -1 & 0 \\

        \hline
      
        0 & \td & 0 & 0 & 0 & 0 & 0 & 0 & 0 & 0 & 1 & 0 & 0 \\

        0 & \td & 0 & 1 & 0 & 0 & 0 & 0 & 0 & 0 & 0 & 0 & 0 \\
        
        0 & \td & 0 & 0 & 0 & 1 & 0 & 0 & 0 & 0 & 0 & 0 & 0 \\

        0 & \td & 0 & 0 & 0 & 0 & 0 & 0 & 0 & 0 & 0 & 0 & 1 

        \end{array}
        \right), \quad \quad
        q = 
        \left(
        \begin{array}{c}
            \mu_\mathrm{M101} \\
            : \\
            \mu_\mathrm{U9391} \\
            \hline
            \mu_\mathrm{N4258} \\
            M_{H,1}^W \\
            \mu_\mathrm{LMC} \\
            \mu_\mathrm{M31} \\
            b_W \\
            M_B \\
            Z_W \\
            \Delta \mathrm{zp} \\
            5 \log H_0 \\
            \Delta \mu_\mathrm{SMC/LMC}
        \end{array}
        \right).
    \end{equation}
\end{widetext}
Note that we have removed the external constraints pertaining to the MW Cepheids, which are instead fitted for here together with the rest of the distance ladder.
Unfortunately, the updated SNIa covariance matrix of \cite{Murakami:2023xuy} that we use in the present work is currently not publicly available, so we are not allowed to provide the data vectors corresponding to eqs.~\eqref{eq:y_update}--\eqref{eq:Lq_update} as we do for in the \Rbase baseline analysis in Section~\ref{sec:R22}.

\section{Matched-period selection}
\label{sec:PerBinning}
As shown in Section~\ref{sec:PerTrends}, there is a trend in $H_0$ as a function of host Cepheid period. In this section, we extend the analysis and calibrate the distance ladder based on Cepheids from a limited range of periods, imposing the range consistently on both anchor and host Cepheids.
That is, fitting only Cepheids within a certain period bin.
As opposed to the approach in Section~\ref{sec:MatchedPerResamp}, this does not enforce the distribution of the anchor and host periods to be exactly the same, but for a small enough range, both distributions will be roughly uniform and therefore approximately the same.
So, fitting the distance ladder with only Cepheids from a specified range can mitigate the difference in periods between anchors and hosts, to a degree depending on the width of the bin. 

In Section~\ref{sec:BrokenPLR}, we saw indications of two breaks in the PLR. Therefore, we do not allow for bins extending across the breaking points at $\p = 0.465 \, \mathrm{dex}$ and $\p = 0.909 \, \mathrm{dex}$. 

\begin{figure}[t]
    \centering
    
    \includegraphics[width=\linewidth]{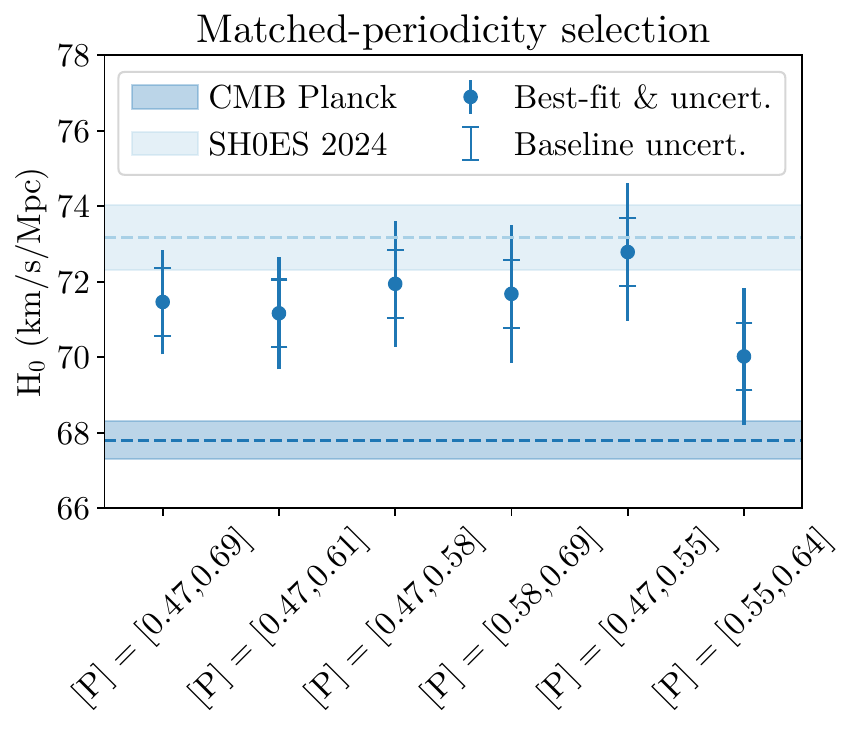}
    
    \caption{Inferred $H_0$ from the Cepheid-based distance ladder, calibrated with Cepheids within a certain period range. The period ranges used for obtaining the different values are recorded along the horizontal axis. We only present the cases where the total uncertainty is less than double the baseline uncertainty of $0.9 \, \mathrm{km/s/Mpc}$. There is a consistent decrease in $H_0$ compared with the SH0ES team. 
    \label{fig:H0binMid}}
\end{figure}

We strive to balance between a small period range (bin), which promotes the conformity between anchor and host periods, and increased uncertainty.
To be systematic, we start by splitting the range $0.465 \, \mathrm{dex} \leq \p \leq 0.909 \, \mathrm{dex}$ into two period bins, and calibrate the distance ladder in each bin. Then, we split the range into three bins and calibrate in each bin, and so on.
Implementing this procedure, there are six period ranges (bins) for which the increase in uncertainty is less than $100 \, \%$, with results shown in in Fig.~\ref{fig:H0binMid}. 
Here, when calibrating the distance ladder, we make a full nonlinear fit of all Cepheids, including the MW.

In Fig.~\ref{fig:H0binMid}, we see a consistent decrease in $H_0$ compared with the reference case, with a shift in the Hubble constant ranging from $\Delta H_0 = -3.1 \, \mathrm{km/s/Mpc}$  to $\Delta H_0 = -0.4 \, \mathrm{km/s/Mpc}$, with a mean shift of $\Delta H_0 = -1.7 \, \mathrm{km/s/Mpc}$. 

Repeating the procedure with Cepheids from the low-period interval $\p < 0.465 \, \mathrm{dex}$, where there is also a significant overlap between anchor and host Cepheids, there is instead a consistent $\simeq + 0.6 \, \mathrm{km/s/Mpc}$ increase in $H_0$.

\end{document}